\documentclass[twocolumn]{aastex63}

\graphicspath{{./}{figures/}}
\usepackage{amsmath}
\usepackage{mathtools}
\usepackage{gensymb}
\usepackage{bm}
\usepackage{esvect}
\usepackage{hyperref}
\DeclareMathOperator*{\argmin}{argmin}
\newcommand{\R}{\mathbb{R}}

\received{November 18, 2020}
\revised{May 27, 2021}
\accepted{May 28, 2021}

\submitjournal{ApJ}

\shorttitle{Surface Imaging of $\lambda$~And}
\shortauthors{Martinez et al.}

\begin{document}

\title{Dynamical Surface Imaging of $\lambda$~Andromedae}

\correspondingauthor{Arturo O. Martinez}
\email{aomartinez@chara.gsu.edu}

\author[0000-0002-3311-4085]{Arturo O.~Martinez}
\affiliation{Center for High Angular Resolution Astronomy, 25 Park Place NE \#605, Atlanta, GA 30303-2911, USA}
\affiliation{Department of Physics and Astronomy, Georgia State University, 25 Park Place NE \#605, Atlanta, GA 30303-2911, USA}

\author{Fabien R.~Baron}
\affiliation{Center for High Angular Resolution Astronomy, 25 Park Place NE \#605, Atlanta, GA 30303-2911, USA}
\affiliation{Department of Physics and Astronomy, Georgia State University, 25 Park Place NE \#605, Atlanta, GA 30303-2911, USA}

\author{John D.~Monnier}
\affiliation{Department of Astronomy, University of Michigan, Ann Arbor, MI  48109-1090, USA}

\author[0000-0002-9288-3482]{Rachael M.~Roettenbacher}
\affiliation{Yale Center for Astronomy and Astrophysics, Department of Physics, Yale University, New Haven, CT 06520-8120, USA}

\author{J.~Robert Parks}
\affiliation{Department of Physics and Astronomy, George Mason University, 4400 University Drive Fairfax, VA 22030-4444, USA}

\begin{abstract}

We present temperature maps of RS CVn star $\lambda$~Andromedae, reconstructed from interferometric data acquired in 2010 and 2011 by the MIRC instrument at the Center for High Angular Resolution Astronomy Array. To constrain the stellar parameters required for this imaging task, we first modeled the star using our GPU-accelerated code SIMTOI. The stellar surface was then imaged using our open source interferometric imaging code ROTIR, in the process further refining the estimation of stellar parameters. We report that the measured angular diameter is $2.742 \pm 0.010$ mas with a limb-darkening coefficient of $0.231 \pm 0.024$. While our images are consistent with those of prior works, we provide updated physical parameters for $\lambda$~Andromedae ($R_{\star}~=~7.78~\pm~0.05~R_{\odot}$, $M_{\star}~=~1.24~\pm~0.72~M_{\odot}$, $\log~L/L_{\odot}~=~1.46~\pm~0.04$).

\end{abstract}


\section{Introduction} \label{sec:intro}

Observing stellar surfaces provides insight to the physics within stellar interiors. We know that stars ranging from pre-main sequence to giants exhibit magnetic spot activity on their surfaces \citep{strassmeier:2009}. Since the advent of space missions, such as CoRoT \citep{baglin:2006b, baglin:2006a} and {\it Kepler} \citep{boruki:2010, koch:2010}, many more stars have been observed to exhibit magnetic activity \citep{frasca:2011,frohlich:2012,roettenbacher:2013,roettenbacher:2016b,nielsen:2019,santos:2019}. These stellar features constitute major sources of uncertainty trying to calculate accurate stellar physical parameters \citep[e.g., $T_{\mathrm{eff}}$ and $R_{\star};$][]{somers:2015}. Starspots have other astrophysical significance tying them to accurately determining exoplanetary parameters. Any uncertainties found in the host star's physical parameters are amplified to any of their planetary parameters, as deriving exoplanetary parameters are dependent on the parent star.

RS Canum Venaticorum (RS CVn) variables are known to present large magnetic starspots \citep{hall:1976, kovari:2015, roettenbacher:2016a, roettenbacher:2017}. These variables are often found in a binary system and the pair often consists of an evolved giant primary with the secondary being a smaller main-sequence companion. Magnetic spots in these systems are often easier to observe because of their relative size to the star, thus making RS CVn variables ideal targets. There are three main techniques routinely employed to image these systems: light-curve inversion, Doppler imaging, and interferometric imaging.

Photometric monitoring of these systems provides straightforward evidence for stellar spots, as shown in many systems observed by the {\it Kepler} spacecraft \citep[e.g.,][]{frasca:2011,frohlich:2012,roettenbacher:2013,roettenbacher:2016b}. The inverse problem of imaging the stellar surface from photometry is called light-curve inversion \citep{wild:1989,roettenbacher:2013}. A main drawback of broadband light-curve inversion is that photometry only provides relative information about the latitude of starspots \citep{harmon:2000} and relies on a prior knowledge of the stellar limb-darkening. Light-curve inversion from multi-band photometry alleviates the latitude ambiguities, hence resulting in more accurate solutions \citep{harmon:2000}.

Doppler imaging \citep{goncharskii:1977,rice:1981} is the class of inverse methods for imaging stellar surfaces from spectroscopic data. This technique uses perturbations of absorption features on a star to better estimate the spot's latitude and longitude. However, there are still uncertainties in determining spot location for stars near edge-on rotation. High-resolution spectra are needed in Doppler imaging to distinguish the features due to the starspots in the absorption lines and to be able to accurately detect their locations. High rotational velocities rotationally broaden absorption lines and are required to ensure that the spectroscopic impact of a spot moving across the surface is shorter than the spot's evolution timescale. \cite{piskunov_wehlau:1990} determined lower bounds enabling Doppler imaging to be from 6 km/s to 15 km/s, which corresponds to spectrograph resolving powers of at least 20,000 to 50,000.

Contrary to Doppler imaging or light-curve inversion, interferometry provides unambiguous evidence that a spot is being shown without any assumptions on latitude. Interferometric modeling allows the determination of angular parameters, such as the inclination or position angle of a spotted star. However, interferometric observations can only be managed on a limited number of targets (i.e., relatively bright targets) when compared to photometric and spectroscopic targets, and furthermore only targets of sufficient angular size can be resolved from Earth. It was only in 2007 that interferometric synthesis imaging became possible \citep{monnier:2007} thanks to longer baselines and the combination of light from four (and now up to six) different telescopes.

The Center for High Angular Resolution Astronomy (CHARA) Array is an interferometric array with six 1 m telescopes, in a Y-shaped configuration, and the world's longest operational baseline (at 330 meters) in optical interferometry. CHARA data has been analyzed to provide detailed images of rapid rotators \citep{monnier:2007, zhao:2009, che:2011}, binary systems \citep{zhao:2008, kloppenborg:2010, baron:2012, kloppenborg:2015}, and nova eruptions \citep{schaefer:2014}. To date, three RS CVn variable stars have been imaged with CHARA: $\lambda$~Andromedae \citep{parks:2021}, $\zeta$ Andromedae \citep{roettenbacher:2016a}, and $\sigma$ Geminorum \citep{roettenbacher:2017}.

$\lambda$~Andromedae (HD 222107; hereafter $\lambda$~And) is a bright G8III-IV RS CVn variable ($V$ = 3.82, $H$ = 1.40) with spots, and is included in the third edition of the Catalog of Chromospherically Active Binary Stars \citep{eker:2008}. It is a single-lined spectroscopic binary system with a rotation period of 54.07 days for the primary \citep{henry:1995b} and has a companion in asynchronous rotation. \cite{walker:1944} found a nearly circular orbit for the system with an eccentricity of $e = 0.084 \pm 0.014$ and an orbital period of $20.5212 \pm 0.0003$~days. The most recent estimate of the effective temperature and mass for the primary star of $\lambda$~And is $4800\pm 100$K and 1.3$^{+1.0}_{-0.6} M_{\odot}$ \citep{drake:2011} with its companion most likely being a low mass main sequence star or a massive brown dwarf based on the mass ratio calculation of $q = 0.12^{+0.07}_{-0.04}$ \citep{donati:1995}. \cite{parks:2021} was the first to do 2D snapshot interferometric imaging of $\lambda$ And using data obtained with CHARA. Their study estimated the angular diameter for the primary of $\lambda$ And to be $2.759 \pm 0.050$ mas, corresponding to a physical radius of $7.831^{+0.067}_{-0.065} R_{\odot}$ given the \textit{Hipparcos} distance of $26.41 \pm 0.15$~pc \citep{vanleeuwen:2007}.

In this paper, we describe the process we followed to obtain a temperature map of the surface of $\lambda$~And. In section \ref{sec:observe}, we present data acquisition and reduction. In section \ref{sec:simtoi}, we describe how we used the interferometric modeling code \texttt{SIMTOI} to obtain initial guesses of stellar parameters. We introduce the \texttt{ROTIR} imaging code in section \ref{sec:rotir}, then its application to the imaging of $\lambda$ And in section \ref{sec:rotir_lamAnd}. The imaging results are compared with previous works in section \ref{sec:compare_previous}. We go on to discuss prospects beyond solid body rotation in section \ref{sec:solid_rot} and the search for the companion of $\lambda$~And in section \ref{sec:binary_search}. Finally, we discuss our conclusions and future work in section \ref{sec:conclusion}.

\begin{deluxetable*}{ccccccl}[t!]
\tablenum{1}
\tablecaption{CHARA Array observations \label{tab:observe}}
\tablecolumns{7}
\tablewidth{0pt}
\tablehead{
\colhead{UT date} &
\colhead{Average} &
\colhead{Baselines} &
\colhead{Number of} &
\colhead{Number of} &
\colhead{Rotation Phase} &
\colhead{Calibrators} \vspace{-5pt} \\
\colhead{} &
\colhead{MJD} &
\colhead{} &
\colhead{$|V|^2$ points} &
\colhead{Closure Phases} &
\colhead{of Primary} &
\colhead{}
}
\startdata
2010 Aug 02 & 55410.4 & S1-E1-W1-W2 & 167 & 88 & 0.0 & 7 And, 37 And \\
 & & S2-E2-W1-W2 & & \\
2010 Aug 03 & 55411.3 & S1-E1-W1-W2 & 454 & 264 & 0.012 & $\sigma$ Cyg, 7 And, 37 And \\
 & & S2-E2-W1-W2 & & \\
2010 Aug 10 & 55418.3 & S1-E1-W1-W2 & 425 & 288 & 0.146 & $\sigma$ Cyg, 7 And, 37 And \\
 & & S2-E2-W1-W2 & & \\
2010 Aug 11 & 55419.3 & S1-E1-W1-W2 & 215 & 136 & 0.164 & $\sigma$ Cyg, 7 And, 37 And \\
2010 Aug 18 & 55426.3 & S1-E1-W1-W2 & 429 & 272 & 0.293 & $\sigma$ Cyg, 7 And, 37 And \\
 & & S2-E2-W1-W2 & & \\
2010 Aug 19 & 55427.3 & S1-E1-W1-W2 & 406 & 264 & 0.312 & $\sigma$ Cyg, 7 And, 37 And \\
 & & S2-E2-W1-W2 & & \\
2010 Aug 24 & 55432.3 & S1-E1-W1-W2 & 526 & 320 & 0.404 & $\sigma$ Cyg, 7 And, 37 And \\
 & & S2-E2-W1-W2 & & \\
2010 Aug 25 & 55433.3 & S2-E2-W1-W2 & 120 & 72 & 0.423 & $\sigma$ Cyg, 7 And, 37 And \\
2010 Sep 02 & 55441.3 & S1-E1-W1-W2 & 522 & 336 & 0.570 & 7 And, 37 And \\
 & & S2-E2-W1-W2 & & \\
2010 Sep 03 & 55442.3 & S1-E1-W1-W2 & 588 & 352 & 0.589 & 7 And, 37 And \\
 & & S2-E2-W1-W2 & & \\
2010 Sep 10 & 55449.3 & S2-E2-W1-W2 & 336 & 192 & 0.718 & 7 And, 37 And \\
2011 Sep 02 & 55806.5 & W1-S2-S1-E1-E2-W2 & 360 & 432 & 0.310 & $\sigma$ Cyg, 7 And, 22 And, HR 653 \\ 
2011 Sep 06 & 55810.5 & W1-S2-S1-E1-E2-W2 & 392 & 376 & 0.384 & $\sigma$ Cyg, 7 And, 22 And, HR 653 \\ 
2011 Sep 10 & 55814.5 & W1-S2-S1-E1-E2-W2 & 360 & 432 & 0.458 & 7 And, 22 And \\ 
2011 Sep 14 & 55818.5 & W1-S2-S1-E1-E2-W2 & 864 & 1104 & 0.532 & 7 And, 22 And, HR 653 \\ 
2011 Sep 19 & 55823.5 & W1-S2-S1-E1-E2-W2 & 808 & 1120 & 0.624 & 7 And, 22 And, HR 653 \\
2011 Sep 24 & 55828.5 & W1-S2-S1-E1-E2-W2 & 200 & 240 & 0.716 & 7 And, 22 And, HR 653, $\eta$ Aur \\
\enddata
\tablecomments{Here we list the UT date, the average modified Julian date of the night of observation, the baselines used in their corresponding configuration, the number of useful squared visibility points obtained for the night, the number of useful closure phase points obtained for the night, the rotation phase for the primary star in $\lambda$ And, and the calibrator stars that were used for each corresponding night. The rotation phase is derived by using the first observation in 2010 as the zero point.}
\end{deluxetable*}
\begin{deluxetable*}{lCcc}[t!]
\tablenum{2}
\tablecaption{Calibrators for $\lambda$ Andromedae \label{tab:cals}}
\tablecolumns{4}
\tablewidth{0pt}
\tablehead{
\colhead{Calibrator Name} &
\colhead{Calibrator Size} &
\colhead{Source} &
\colhead{Epoch Used} \vspace{-5pt} \\
\colhead{} &
\colhead{(mas)} &
\colhead{} &
\colhead{}
}
\startdata
7 And (HD 219080) & 0.65 \pm 0.03 & \cite{mourard:2015} & 2010 \\
37 And (HD 5448) & 46.66 \pm 0.06 & \cite{roettenbacher:2016a}\tablenotemark{a} & 2010 \\
$\sigma$ Cyg (HD 202850) & 0.542 \pm 0.021 & \cite{zhao:2008} & 2010 \\
\hline
7 And (HD 219080) & 0.676 \pm 0.047 & SearchCal \citep{bonneau:2006} & 2011 \\
$\sigma$ Cyg (HD 202850) & 0.54 \pm 0.02 & \cite{barnes:1978} & 2011 \\
22 And (HD 571) & 0.591 \pm 0.041 & SearchCal \citep{bonneau:2006} & 2011 \\
HR 653 (HD 13818) & 0.646 \pm 0.045 & SearchCal \citep{bonneau:2006} & 2011 \\
$\eta$ Aur (HD 32630) & 0.336 \pm 0.023 & SearchCal \citep{bonneau:2006} & 2011 \\
\enddata
\tablecomments{The angular sizes for the 2011 epochs are based on what was reported from \cite{parks:2021} since we use their reduced and calibrated data. We use updated angular sizes for each calibrator star in the 2010 epoch since we do a new and separate reduction and calibration. The differences in angular sizes for 7~And and $\sigma$ Cyg used between the two years are small and within their 1$\sigma$ errors.
\tablenotetext{a}{This is the semi-major axis angular separation of the binary calculated by \cite{roettenbacher:2016a}.}}
\end{deluxetable*}

\section{Observations} \label{sec:observe}
We reuse the 2010 and 2011 data from \cite{parks:2021}, shown in Table \ref{tab:observe} and calibrators in Table \ref{tab:cals} used for each respective year, for our analysis. These data were obtained using the CHARA Array \citep{tenbrummelaar:2005} with the Michigan Infra-Red Combiner \citep[MIRC;][]{monnier:2004} in H-band with the median wavelength of 1.65 $\micron$. The observations were done in prism mode (R = 50) which contain eight spectral channels. The data taken in 2010 were taken with a combination of four out of six telescopes which provide six visibilities, three independent bispectrum amplitudes (triple amplitudes), and three independent bispectrum phases (closure phases). The 2011 data set benefited from MIRC having been upgraded earlier that year, allowing for simultaneous use of all six telescopes. These upgrades provided data sets to acquire up to 15 visibilities, 10 independent triple amplitudes, and 10 independent closure phases for each spectral channel.

\subsection{Data Reduction} \label{sec:interferometry_reduction}

\cite{parks:2021} detail the reduction steps and error corrections but we will briefly note some of their steps here. The data were reduced using the official IDL pipeline for reducing MIRC data \citep{monnier:2007}. Each block of raw fringe data contained coadded frames, and were corrected for any instrumental effects by background subtraction in order remove instrumental noise and foreground normalization to correct for any pixel-to-pixel variation. Raw square visibilities, closure phases, and triple amplitudes are output through the use of Fourier transforms and are photometrically calibrated. The data were corrected for the atmospheric coherence time and optical changes in the beam path with the use of calibrator stars that were taken either immediately before or after the target $\lambda$~And. 

In the 2010 data, one of the calibrators 37~And (HD~5448) was found to be a binary by \cite{che:2012} and had its orbit fully characterized by \cite{roettenbacher:2016a}. \cite{parks:2021} formed a comparison of using 37~And as either a single star calibrator or as a binary calibrator. They found that these comparisons only incurred an error of 1.24\% for the square visibilities, which is well below the multiplicative error correction, and a closure phase standard deviation of 1.14$^{\circ}$. We execute a separate reduction and calibration for the 2010 data set using the official MIRC reduction pipeline in order to correct for the 37 And binary calibrator. We use the more recent calibrator diameter estimates, whose values differ from \cite{parks:2021}, for this new reduction and calibration. The data uncertainties also go through a post-calibration process to account for known systematic errors of the MIRC instrument.

\begin{figure*}[ht]
    \centering
    \includegraphics[width=0.8\textwidth]{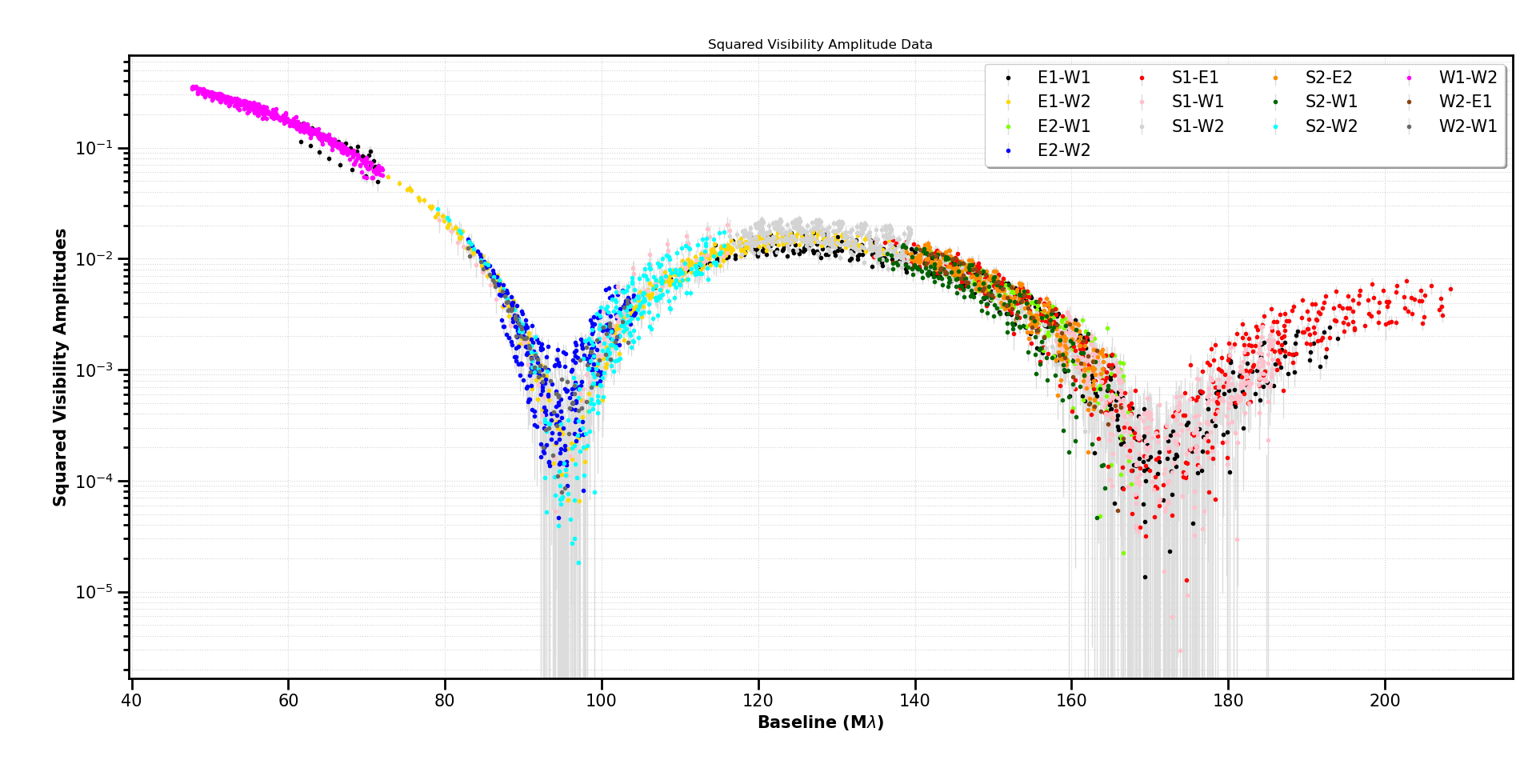}
    \includegraphics[width=0.8\textwidth]{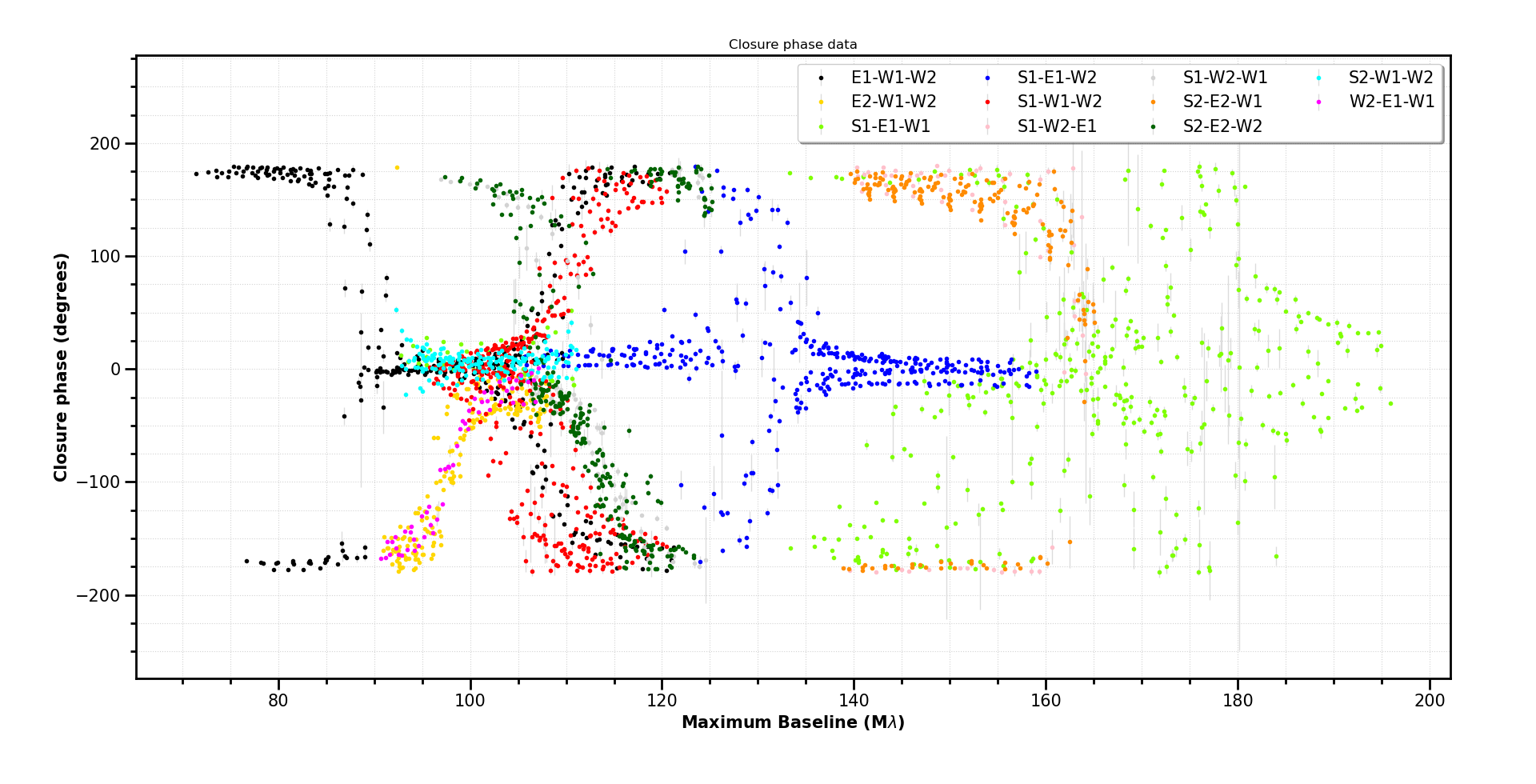}
    \caption{Top: $\vert$V$\vert^2$ points are plotted against the baseline length (in M$\lambda$) for a given baseline pair for all the data of $\lambda$~And from the 2010 epoch. Bottom: Closure phase points are plotted against the baseline length (in M$\lambda$) for the given baseline trio for the same 2010 data.
    \label{fig:v2_2010lamAnd}}
\end{figure*}

For the 2010 data we kept the same systematic errors as \cite{parks:2021}. These errors are different compared to the 2011 data set as the quality of the 2010 data are taken with a four telescope configuration and are of lower quality while the higher quality 2011 data are taken with a six telescope configuration. A 15\% multiplicative error correction was used in association with the transfer function, a 2 $\times$ 10$^{-4}$ additive error correction was used in association with bias at low amplitudes for the square visibilities, and a 20\% multiplicative error correction and a 1 $\times$ 10$^{-5}$ additive error correction was used for the triple amplitudes. The same 1$^{\circ}$ error floor was used for the closure phases as was used in \cite{zhao:2011}. We present the square visibilities and closure phases for the 2010 data set in Figure \ref{fig:v2_2010lamAnd}.

\begin{figure*}[ht]
    \centering
    \includegraphics[width=0.8\textwidth]{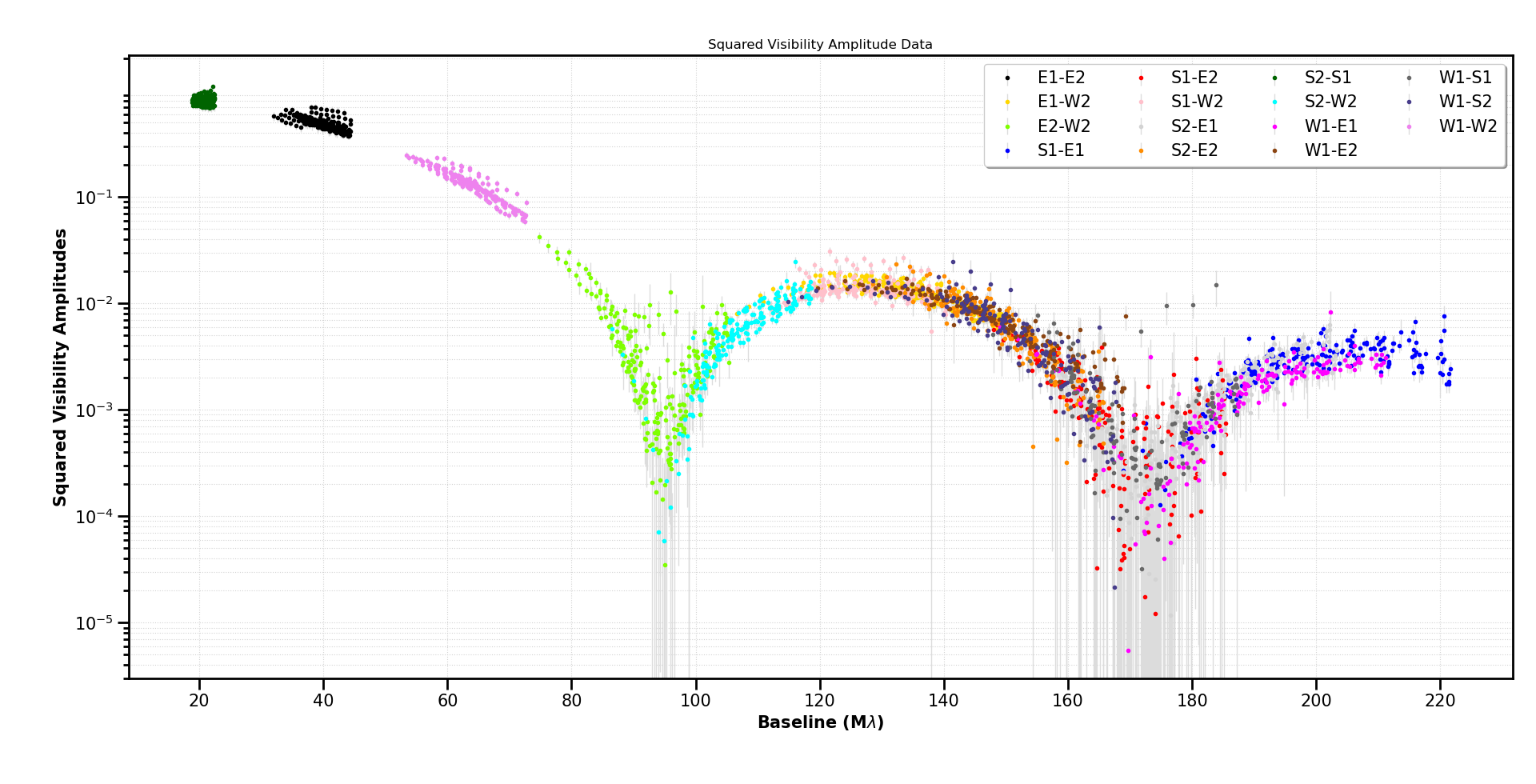}
    \includegraphics[width=0.8\textwidth]{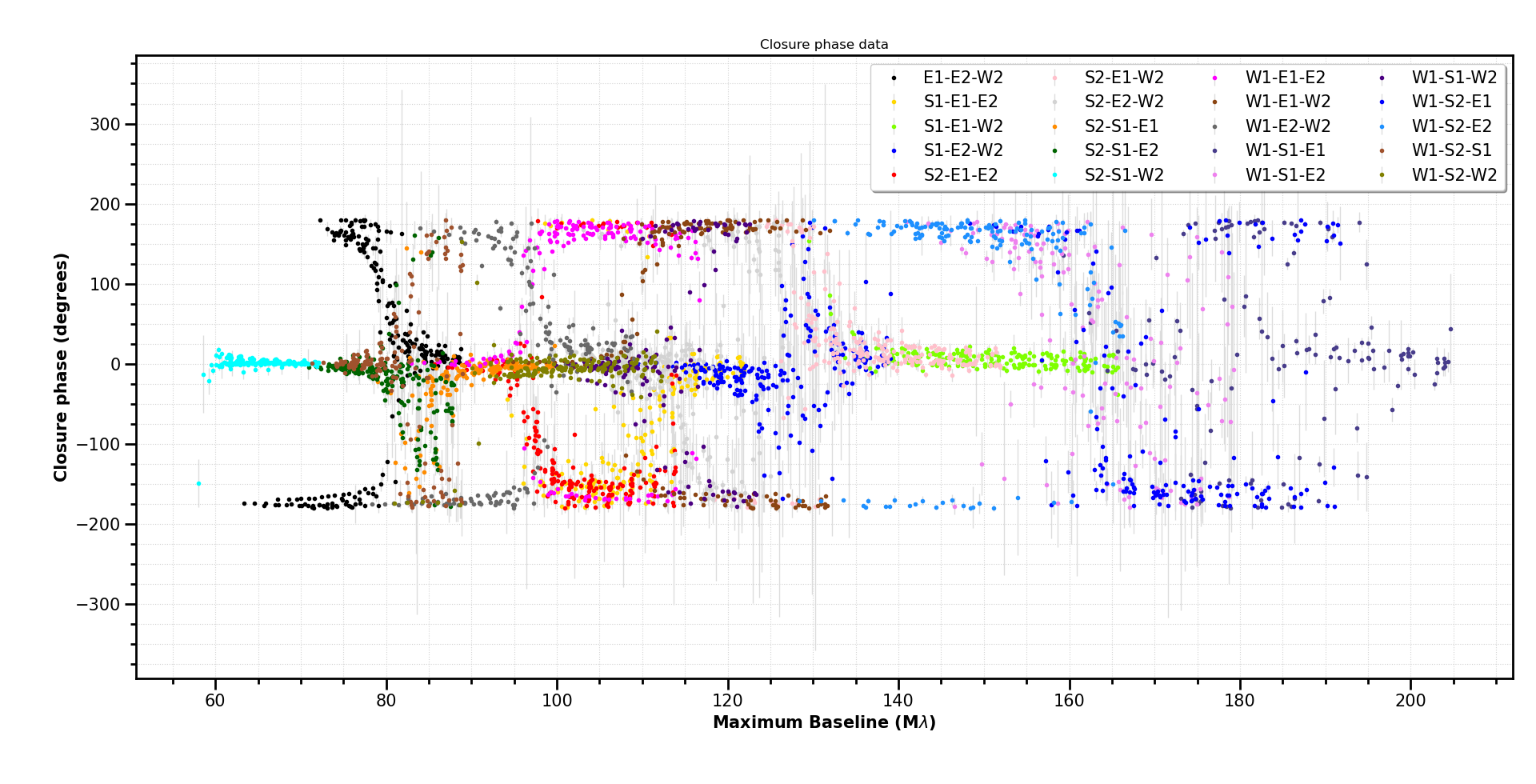}
    \caption{Top: $\vert$V$\vert^2$ points are plotted against the baseline length (in M$\lambda$) for a given baseline pair for all the data of $\lambda$~And from the 2011 epoch. Bottom: Closure phase points are plotted against the baseline length (in M$\lambda$) for the given baseline trio for the same 2011 data.
    \label{fig:v2_2011lamAnd}}
\end{figure*}

We use the same calibrator diameter estimates listed in \cite{parks:2021} since the 2011 data set has been reduced and calibrated. Even though different angular sizes were used for the calibration of the 2010 and 2011 data set for 7~And and $\sigma$~Cyg, the differences between the two angular sizes reported in Table \ref{tab:cals} are small and within their respective 1$\sigma$ errors. Systematic errors were taken into account during calibration similar to that of \cite{monnier:2012}. A 10\% multiplicative error correction was used in association with the transfer function for the 2011 data and a 2 $\times$ 10$^{-4}$ additive error correction was used for the square visibilities. A 15\% multiplicative error correction was used and a 1 $\times$ 10$^{-5}$ additive error correction was used for all the triple amplitude data. Lastly, the same 1$^{\circ}$ error floor was used for the closure phases just as it was presented in \cite{zhao:2011}. We present the square visibilities and closure phases for all of the 2011 data set in Figure \ref{fig:v2_2011lamAnd}.

\section{Modeling \texorpdfstring{$\lambda$}{lambda}~And with SIMTOI} \label{sec:simtoi}
The SImulation and Modeling Tool for Optical Inteferometry (\texttt{SIMTOI}) is an interferometric modeling code\footnote{\url{https://github.com/bkloppenborg/simtoi}} \citep{kloppenborg_baron:2012a, kloppenborg_baron:2012b, kloppenborg:2015} that uses a Graphical Processor Unit (GPU) to represent stars and their environments in a three-dimensional framework. In \texttt{SIMTOI}, the stellar intensity maps are two-dimensional textures applied on top of orbiting/rotating three-dimensional stars. Once the scene is rendered, the GPU also powers the fast computation of interferometric observables. \texttt{SIMTOI} offers a large choice of global and local optimizers to solve {\it Maximum A Posteriori} (MAP) or model selection problems. Our first goal in using \texttt{SIMTOI} was to derive initial guesses for $\lambda$~And's stellar parameters (such as its rotation axis), since our imaging code would be too slow to wade through the entire parameter space. Our second goal was to assess the potential number of spots present on the star via model selection. Both tasks were solved using the MultiNest optimizer \citep{feroz_hobson:2008, feroz:2009, feroz:2019}, which implements the Importance Nested Sampling algorithm.

\begin{deluxetable*}{ccccc}
\tablenum{3}
\tablecaption{\texttt{SIMTOI} Model Results \label{tab:simtoi_results}}
\tablecolumns{5}
\tablewidth{0pt}
\tablehead{
\colhead{Number of} &
\multicolumn{2}{c}{Importance Nested} &
\multicolumn{2}{c}{$\chi_{\nu}^2$} \vspace{-5pt} \\
\colhead{spots} &
\multicolumn{2}{c}{Sampling value (ln $Z$)} &
\multicolumn{2}{c}{} \\
\cline{2-5} \vspace{-15pt} \\
& \colhead{2010 data} & \colhead{2011 data} & \colhead{2010 data} & \colhead{2011 data} \vspace{-15pt} \\
}
\startdata
3 & -67599.448761 & 10433.734767 & 56.855263 & 8.020829 \\ 
4 &  -5968.401012 & 36157.820383 & 17.114268 & 3.102330 \\ 
5 &   1420.128960 & 32785.721919 & 11.372939 & 3.710609 \\
6 & -45295.096429 & 30193.016672 & 38.679240 & 4.211333 \\ 
\hline
Physical parameters & Value & & \\
(4 spot model based on 2011 data) & & & \\
\hline
\multicolumn{1}{l}{$R_{\star}$ (mas)} & 1.37 & & \\ 
\multicolumn{1}{l}{Limb-darkening coefficient} & 0.22 & & \\
\multicolumn{1}{l}{Inclination (deg)} & 86.4 & & \\
\multicolumn{1}{l}{Position Angle (deg)} & 26.7 & & \\
\multicolumn{1}{l}{Rotation Period (days)} & 54.2 & & \\ 
\enddata
\tablecomments{Higher ln $Z$ value is better, lower $\chi_{\nu}^2$ is better. No error bars are calculated since the models from \texttt{SIMTOI} using Multinest does not currently generate reliable error bars. We rely on the imaging results for more precise measurements and calculation of errors.}
\end{deluxetable*}
\subsection{Modeling \texorpdfstring{$\lambda$}{lambda}~And}

We devised models of $\lambda$~And with different number of circular spots, from three to six. Six parameters were used to model the star itself: rotation period, rotation axis (inclination and position angles), temperature, angular diameter, and coefficient of the power limb-darkening law \citep{hestroffer:1997}. The stellar parameters were given uniform prior distributions within a wide range of values, based on the stellar parameters listed in \cite{parks:2021} as a starting point (e.g. $\pm 20\degree$ for angular parameters). Four parameters were used per spot: longitude, latitude, diameter, and flux. These spot parameters were also given uniform distribution. In particular, their location was not constrained.

For each data set -- 2010 or 2011 -- SIMTOI renders an image per epoch (day). The rendering resolution was set to a 64 $\times$ 64 image with a 0.05 mas/pixel resolution. MultiNest was run for each model and converged after a few hours, providing MAP parameter values, as well as the marginal likelihood values (the so-called $\log Z$).

\subsection{Modeling results}
We report the $\chi^2$ and $\log Z$ values for each spot model in Table \ref{tab:simtoi_results}. We also provide the approximate nominal values for the physical parameters. MultiNest does provide error bars, but since they do not account for systematic errors, they are vastly underestimated. While one could bootstrap the data before MultiNest runs, this would be too computationally intensive and yet still imprecise due to our approximate modeling of spots. Our model spots are circular, which may be an unrealistic assumption, but is sufficient to identify the main potential location of intensity peaks on the surface. The $\log Z$ values are maximal for the five spot model for the 2010 data, and the four spot model for the 2011 data. The corresponding reduced $\chi^2$ values are low for the 2011 data and much higher for 2010. Setting aside the possible differences in error calibration between 2010 and 2011, this would indicate that the 2010 surface map is much more complex than the 2011 one (which we did confirm during imaging).

We ultimately choose the 4 spot model for the 2011 data as the best representative model that produces the most accurate parameterization of $\lambda$ And. The estimated 54.2 day rotation period of the primary from our model using the 2011 data set is consistent with other works. \cite{henry:1995b} reports a rotation period 54.07 days from their photometric analysis while \cite{parks:2021} reports a $54.02 \pm 0.88$ day rotation period from their own photometric analysis and an average of a $56.9 \pm 8.8$ day rotation period from their interferometric analysis. While the 2010 data set had a larger rotation phase coverage than the 2011 data set, the rotation period based on the 2011 data are overall more reliable based on MultiNest results and the fitting of the model to the data. This is most likely due to the larger amount of ($u,v$) coverage, number ($u,v$) points, triple amplitudes, and closure phase points in the 2011 data set compared to the 2010 data set. This calculated period from the four spot model using the 2011 data is consistent with previous works.


\section{ROTIR} \label{sec:rotir}
Our code ROTational Image Reconstruction\footnote{\url{https://github.com/fabienbaron/ROTIR.jl}} (\texttt{ROTIR}) is an open source Julia code (\citealp{baron:2018}, \citealp{baron:2021}) which models the stellar surface temperatures of single stars or binary systems as two-dimensional arrays on top of a stellar geometry. The stellar geometry itself is defined either by analytic formulas (ellipsoids, fast rotators) or by solving Roche equations. In imaging and model-fitting problems, ROTIR makes use of the optimization packages \textit{OptimPack} \citep{thiebaut:2002} and NLopt \citep{johnson:2008} to maximize the posterior probability of the model.

\subsection{Geometry setup} \label{sec:geometry}
ROTIR relies on the package \texttt{OITOOLS}\footnote{\url{https://github.com/fabienbaron/OITOOLS.jl}} \citep{baron:2019}  to read in our data, split up or combine our data temporally, and plot any images featured in this work.

Once the interferometric data are read, we define the stellar parameters and orientation of our object. Our code requires several parameters: the angular size at the pole in milliarcseconds, the surface temperature, the fraction of the critical angular velocity if the star is rapidly rotating, the limb-darkening law and its corresponding coefficient(s), the exponent needed if there is any gravity darkening \citep{vonzeipel:1924}, the difference in angular velocity between the equator and the pole, the inclination, position angle, and rotation period of the star. Our code allows the user to choose between three different limb-darkening laws: a quadratic law, logarithmic law, or Hestroffer law \citep[commonly known as the power law;][]{hestroffer:1997}. 

Our geometrical setup starts with selecting a tessellation scheme. Two schemes have been implemented so far: the HEALPix tessellation \citep{gorski:2005} and the latitude/longitudinal scheme. HEALPix presents the advantage of equal area tessels, provided the star does not depart too much from a spherical shape. The latitude/longitudinal scheme allows for simulating differential rotation, but requires more tessels to represent the surface. As part of this work we tested both tessellation schemes, which result in qualitatively identical maps. Most results presented in this paper were obtained with the latitude/longitude scheme. The number of pixels per angular diameter was chosen based on the estimated angular diameter size divided by the imaging resolution limit. Therefore, the minimum total number of pixels required across the surface of a star would simply be the number of latitude pixels times the number longitude pixels.

For the latitude/longitude scheme, the number of latitude pixels is based on the number of pixels per angular diameter since the latitude range spans from $-90^\circ$ to $90^\circ$ and number of longitude pixels is twice the number of pixels per angular diameter since the longitude ranges from $0^\circ$ to $360^\circ$. We number the vertices of the polygon by $1, 2, 3, 4$ in a counterclockwise direction when viewed along the direction of the normal. A fifth element is also included for each pixel and defined to be at the center of each pixel.

Once the user has chosen a tessellation scheme and calculated the number of pixels required for imaging, the user then has the choice of choosing between three different geometries: a scaled unit-sphere, an oblate spheroid, or a Roche object\footnote{Technically, the model of the star is a polyhedron since the surface is made up of many different pixels and not one solid surface. In order to describe the overall shape of the star, we choose to name them as 3D objects instead of polyhedrons.}. The order in which the pixels are mapped out on the surface of the star are counterclockwise when viewed along normal of the positive $z$ direction on the ($x,y,z$) plane.

The surface area $\mathbf{A}_{n}$ is calculated for all $n$ pixels in order to determine the amount of relative flux coming from the star with the following 
\begin{equation}
     \mathbf{A}_{n} = \frac{1}{2} \sum_{j=1}^{m} (\mathbf{v}_{j}\wedge\mathbf{v}_{j+1}) \cdot \hat{\mathbf{z}}
\end{equation}
where $\mathbf{v}$ is the vector of ($x,y$) projected positions of the $n^{\mathrm{th}}$ pixel in a 2-dimensional ($x,y$) plane at the $j^{\mathrm{th}}$ corner with $m$ number of corners in the polygon of choice, $\cdot$ is the scalar product, and $\wedge$ the vector cross product operator. The $m+1$ corner here points back to the first corner of the pixel.

Once the surface area of each pixel is calculated with the desired limb-darkening law, the Fourier transform $\mathbf{S}$ is done on every pixel for a 3-dimensional object \citep{lee_mittra:1983,chu_huang:1989,mcinturff:1991} in order to compare the frequencies of our data on the ($u,v$) plane by using the following equation
\begin{align}
    \begin{split}
     \mathbf{S}(\mathbf{k}) =\sum_{j=1}^{m} & \hat{\mathbf{z}} \cdot \left[\left(\mathbf{v}_{j+1} - \mathbf{v}_{j} \right)\wedge \mathbf{k} \right] 
    \frac{\mathrm{sinc}[\mathbf{k} \cdot \left(\mathbf{v}_{j+1} + \mathbf{v}_{j} \right)]} {i2\pi \vert \mathbf{k} \vert ^2}
     \\
     &\times \mathrm{exp}[-i\pi \mathbf{k} \cdot \left(\mathbf{v}_{j+1} + \mathbf{v}_{j} \right)] 
    \end{split}
\end{align} 
where $\mathbf{k}$ is a vector containing each $u$ and $v$ frequency on the ($u,v$) Fourier plane.
We use the flux to visibility matrix $\mathbf{S}$ to compute the model visibilities using:
\begin{eqnarray}
\mathbf{V} = \frac{\mathbf{S  (L\circ T)}}{\mathbf{A}^{\top} (\mathbf{L}\circ \mathbf{T}) }
\end{eqnarray}
where $\mathbf{V}$ is the model complex visibility vector, $\mathbf{T}$ is the temperature map vector, $\mathbf{L}$ is the limb-darkening map, $\circ$ is the Hadamard (element by element) vector product, and the division is the Hadamard division.

\subsection{Differential rotation option}
The user can select whether or not to turn on the option to simulate differential rotation. The equation for differential rotation \citep{henry:1995b} used in our code is in the form
\begin{equation}
    {\Omega(\Psi) = \Omega_{eq} - \Delta \Omega \sin^{2}{\Psi}}
\end{equation}
where $\Psi$ is the latitude, $\Omega$($\Psi$) is the rotation rate at a specific latitude, $\Omega_{eq}$ is the rotation rate at the equator, and $\Delta \Omega$ is the difference in angular velocity between the equator and the pole. This difference between angular velocity in the equator and the pole is related to the differential rotation coefficient, $k$, or the surface shear parameter, $\alpha$, commonly found in the literature \citep[e.g.,][]{henry:1995b, davenport:2015, kovari:2015} and  is defined through the following equation
\begin{equation}
    {k = \frac{\Omega_{eq} - \Omega_{pole}}{\Omega_{eq}} ,}
\end{equation}
or in terms of the polar and equatorial rotational periods as
\begin{equation}
    {k = \frac{P_{pole} - P_{eq}}{P_{pole}}.}
\end{equation}

\subsection{First use of regularization}
Fitting a model to the data with no prior constraints will produce unrealistic images due to overfitting. The {\it Maximum A Posteriori} method balances the likelihood term with our prior expectations of what the temperature map should look like. The optimal temperature map $x_{opt}$ is then found as:
\begin{equation} \label{eq:reg_tv}
    {x_{opt} = \argmin_{x \in \R^n} \mathrm{\{} \chi^2 \mathrm{(}x\mathrm{)} + \mu R\mathrm{(}x\mathrm{)} \mathrm{\}}}
\end{equation}
where $\chi^2$($x$) is the chi square fit of the data to the model, $R(x)$ is the regularization, and $\mu$ is the hyperparameter setting the relative weight of the regularizer versus the likelihood.

We implemented three different regularizations for use in \texttt{ROTIR}: positivity, $l_2$ norm, and total variation. Positivity enforces a non-negative temperature map. The $l_2$ norm takes the square root of the sum of square values for each pixel and penalizes pixel values straying too far from the average value. Our third regularizer is total variation which computes the spatial gradient of the model image and penalizes large temperature fluctuations between neighboring pixels such that it shows smoother transitions on a local scale.

\section{Applying ROTIR to \texorpdfstring{$\lambda$}{lambda}~And} \label{sec:rotir_lamAnd}
\begin{figure*}[ht!]
   \begin{center}
   \begin{tabular}{ccc} 
   \includegraphics[width=0.333\textwidth]{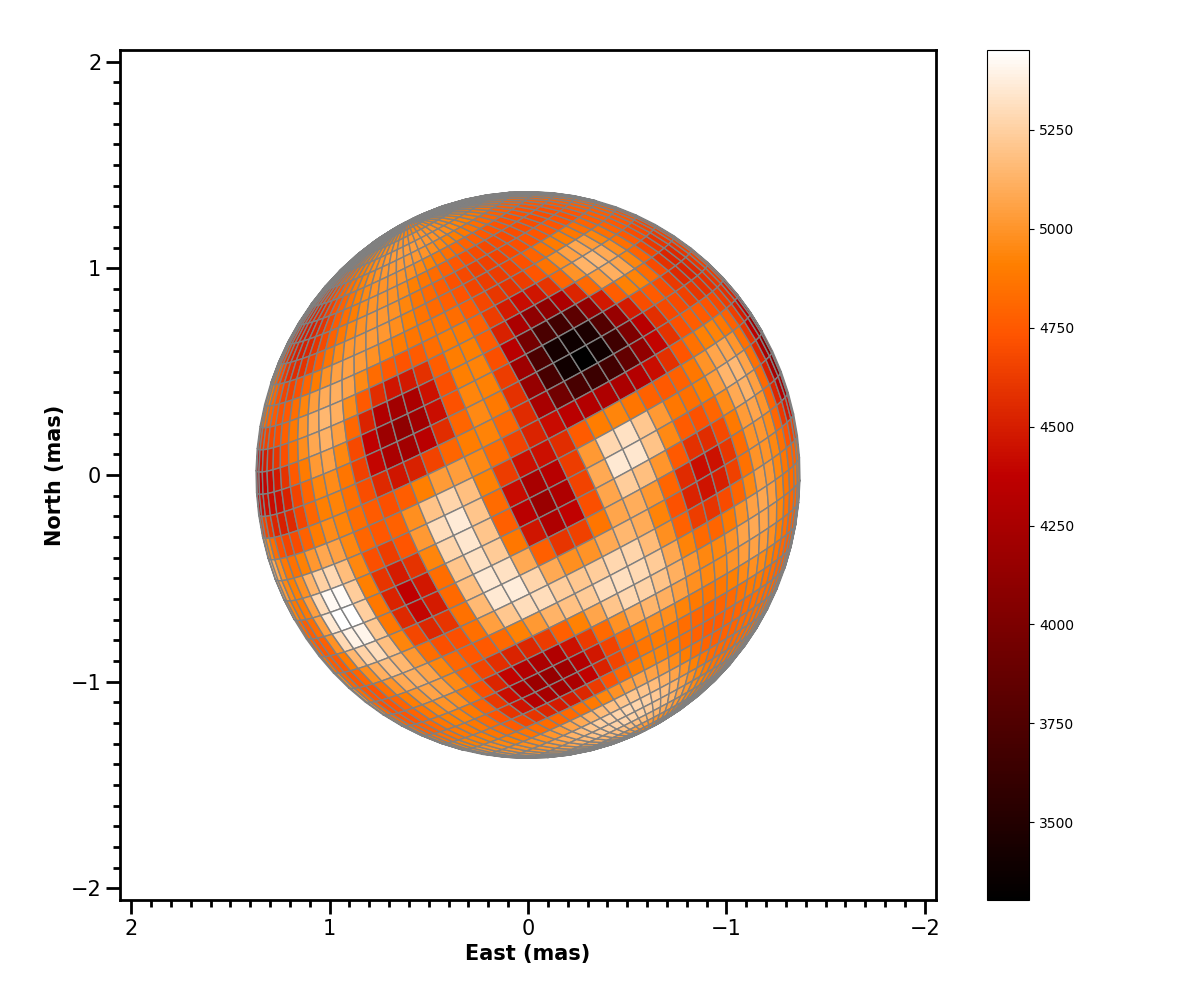}
   \includegraphics[width=0.333\textwidth]{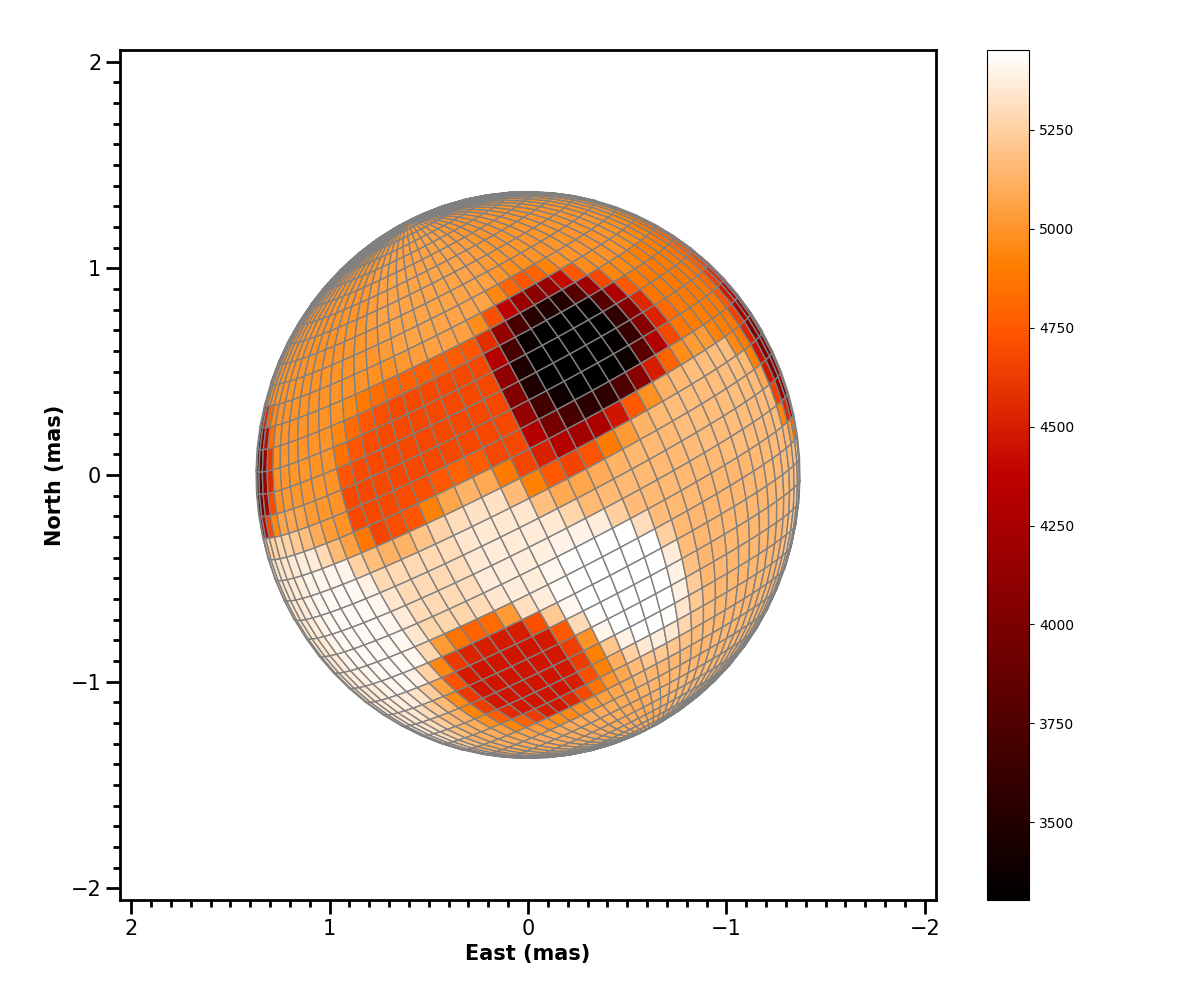}
   \includegraphics[width=0.333\textwidth]{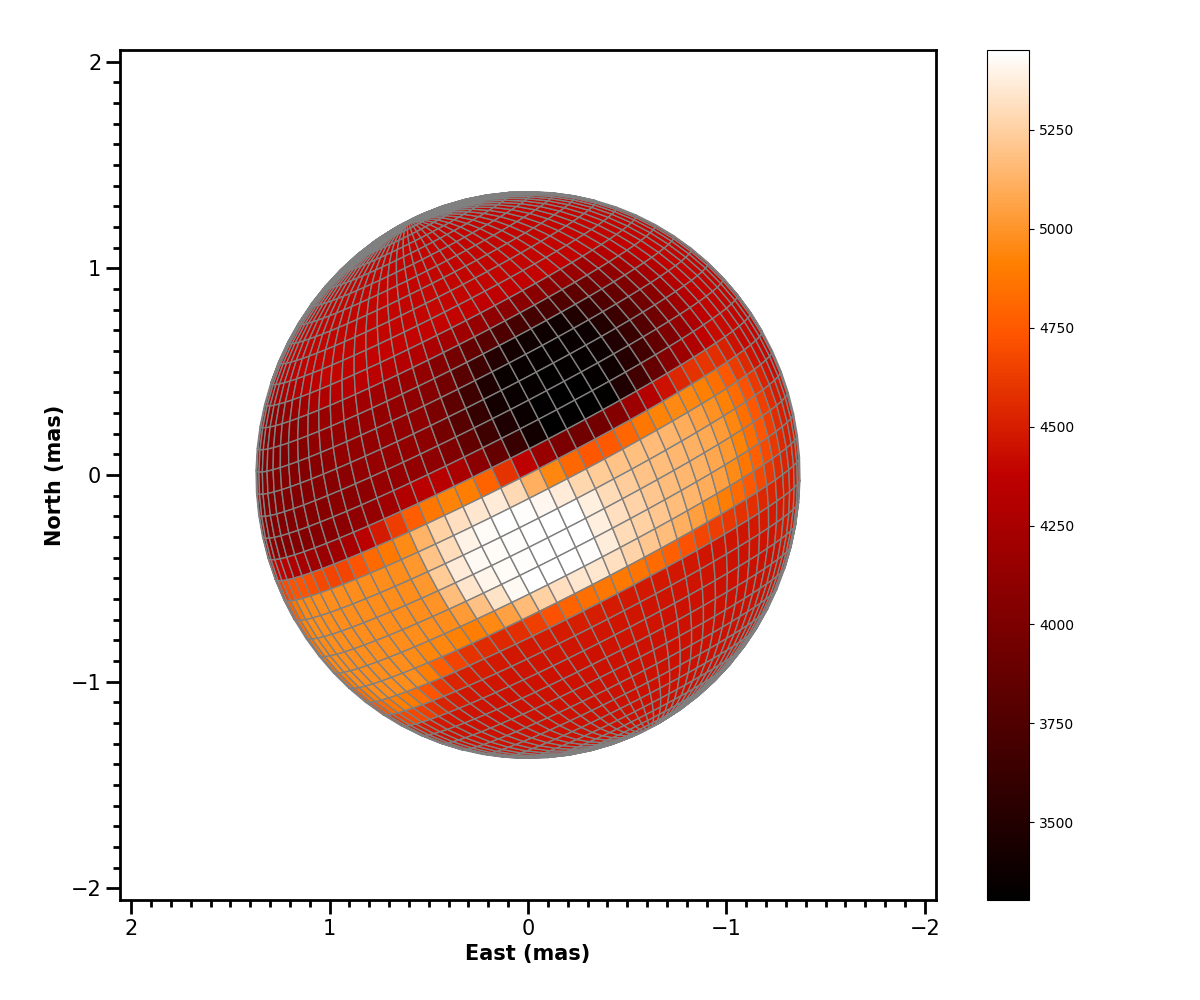}
   \end{tabular}
   \end{center}
   \caption{Left: Here we show an example of a reconstruction made with a very weak hyperparameter ($\mu$ = 0.0001). This is close to the classic example of overfitting an image based on the data. Middle: Here we show where there is a good balance of fitting the data to a model and the use of a hyperparameter ($\mu$ = 0.01). Right: Here we show an example where the hyperparameter is dominant ($\mu$ = 0.5) and very loosely based on the data fitting the model. All three temperature maps are in Kelvin. \label{fig:diff_hyperparam_tempmap}}
\end{figure*}

For $\lambda$~And, we use positivity and total variation as the two regularizers necessary to determine the best image. Using the $l$-curve method \citep{renard:2011}, we choose a weight of $\mu~=~0.01$ that has a small amount of regularization before entering into a regularization dominated regime. We show examples of strong and weak regularization in Figure \ref{fig:diff_hyperparam_tempmap} for data of $\lambda$~And taken on 2011-Sep-14 to prove why we need a good balance between regularization and pure model fitting when finding an optimum image. 

In order to determine of the number of tessels required on the surface of the star, we use the parameters we obtained from modeling $\lambda$~And using \texttt{SIMTOI}. Knowing that the CHARA angular resolution limit is $\theta \approx 0.60$ mas at H-band ($\lambda=1.61 \micron$), we estimate that we need 40 pixels across the visible equator to meet Nyquist sampling (imaging resolution limit is $\theta \approx 0.30$ mas in H-band). Therefore, we use 80 pixels around each latitude, including pixels behind the star, and 40 pixels across each longitude for a total of 3200 pixels on the surface of the star. Our sampling of pixels across the resulting images are solely based on the number of pixels on the surface on the star and not the overall field as the field size can be arbitrarily chosen based on the plotting axes.

\subsection{A first look at imaging} \label{sec:rotir_optimpack}
In order to find the best geometrical setup for primary star in $\lambda$~And, we test both a spherical star and a Roche lobe shape to see if there is any signs of major Roche lobe overflow. While \cite{donati:1995} and \cite{parks:2021} both suggest that there is no Roche lobe overflow, we decide to investigate this for $\lambda$ And since slight oblateness was found in another RS CVn variable, $\zeta$ Andromedae \citep{roettenbacher:2016a}.

We start with the parameters from \texttt{SIMTOI} to create our spherical star and, with the addition of the longitude of the ascending node, argument of periapsis, and eccentricity found in \cite{walker:1944}, create our Roche lobe geometry. \cite{donati:1995} states that $\lambda$ And is coplanar, therefore we use the inclination rotation axis of the primary star as the inclination of the orbit. We use the same hyperparameter and apply a uniform temperature map across the whole star as an initial condition for both geometries. Using a Julia package called \textit{OptimPack}\footnote{\url{https://github.com/emmt/OptimPackNextGen.jl}} that solves for an optimum temperature map through a quasi-Newtonian method \citep{thiebaut:2002}, we obtain for the best temperature map given our all our data in a given year. This algorithm compares the Fourier transforms from Section \ref{sec:geometry} to the 2011 data to solve for the best temperature map.

The resulting criterion for the Roche lobe geometry is higher ($\chi^2$($x$)$ + \mu R$($x$) = 6288) when directly comparing it to a spherical geometry (criterion = 4489). We also find that the pole-to-equator ratio at the L1 Lagrangian point for the primary is 0.9967. With these two calculations, we determine that a spherical geometrical shape for the primary of $\lambda$~And is an acceptable approximation for the true shape of the star.

\begin{figure*}[ht!]
    \includegraphics[width=0.5\textwidth]{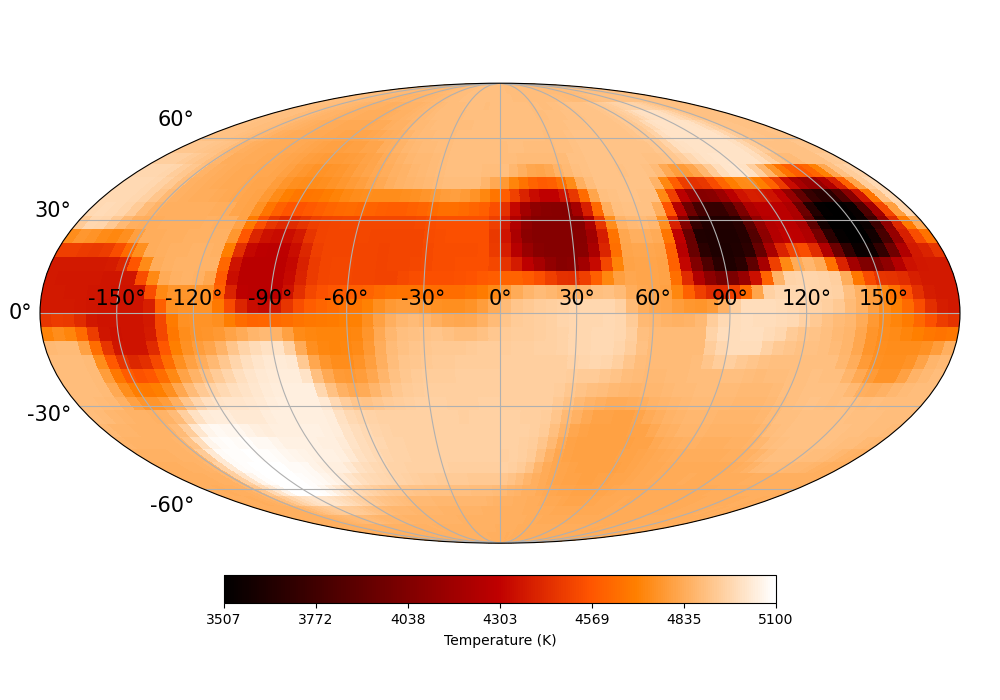}
    \includegraphics[width=0.5\textwidth]{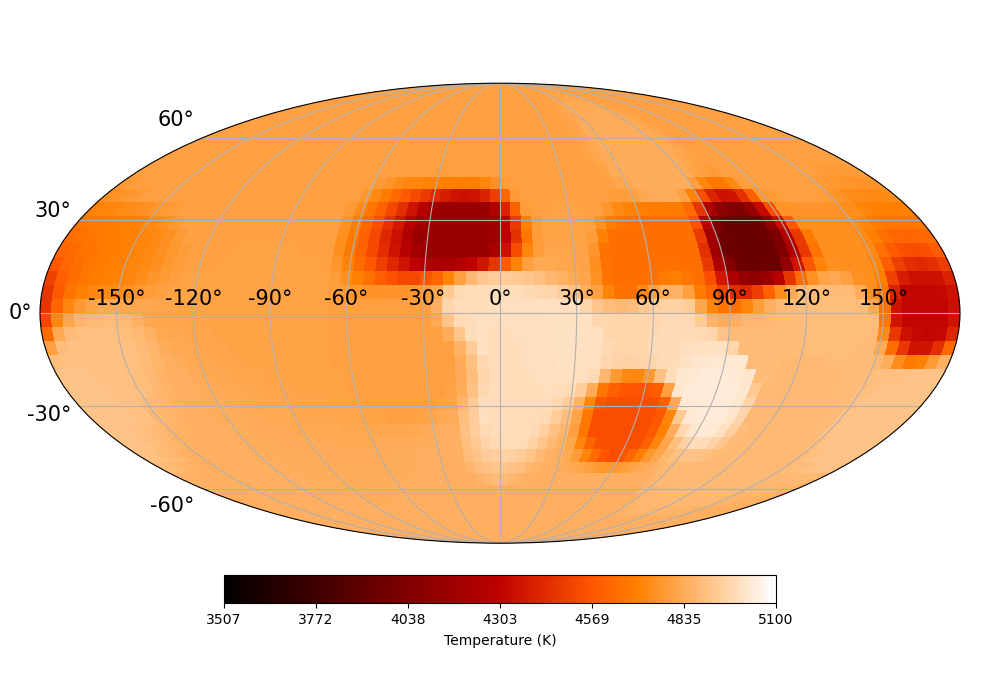}
    \caption{Here we show a Mollweide plot of $\lambda$~Andromedae for the 2010 epoch (left) and 2011 epoch (right) using our \texttt{ROTIR} code. We combine our 11 nights of data in 2010 across 39 nights and 6 nights of data in 2011 across 22 nights to make the temperature map for the 2010 epoch and 2011 epoch, respectively. Both plots use the first date of the 2010 data as the zero point rotation phase and are shifted accordingly. We note that the pixels not within the observing line-of-sight are calculated by starting at the effective temperature from \cite{drake:2011} and modified through \textit{OptimPack}. Areas where the $\lambda$ And is not observed (i.e., latitudes below -85$^\circ$ for both epochs and longitudes between -124$^\circ$ and -90$^\circ$ for the 2011 epoch) have temperatures near the input effective temperature of 4800 K.
    \label{fig:Mollweide_map}}
\end{figure*}

\begin{figure*}[ht!]
    \includegraphics[width=0.5\textwidth]{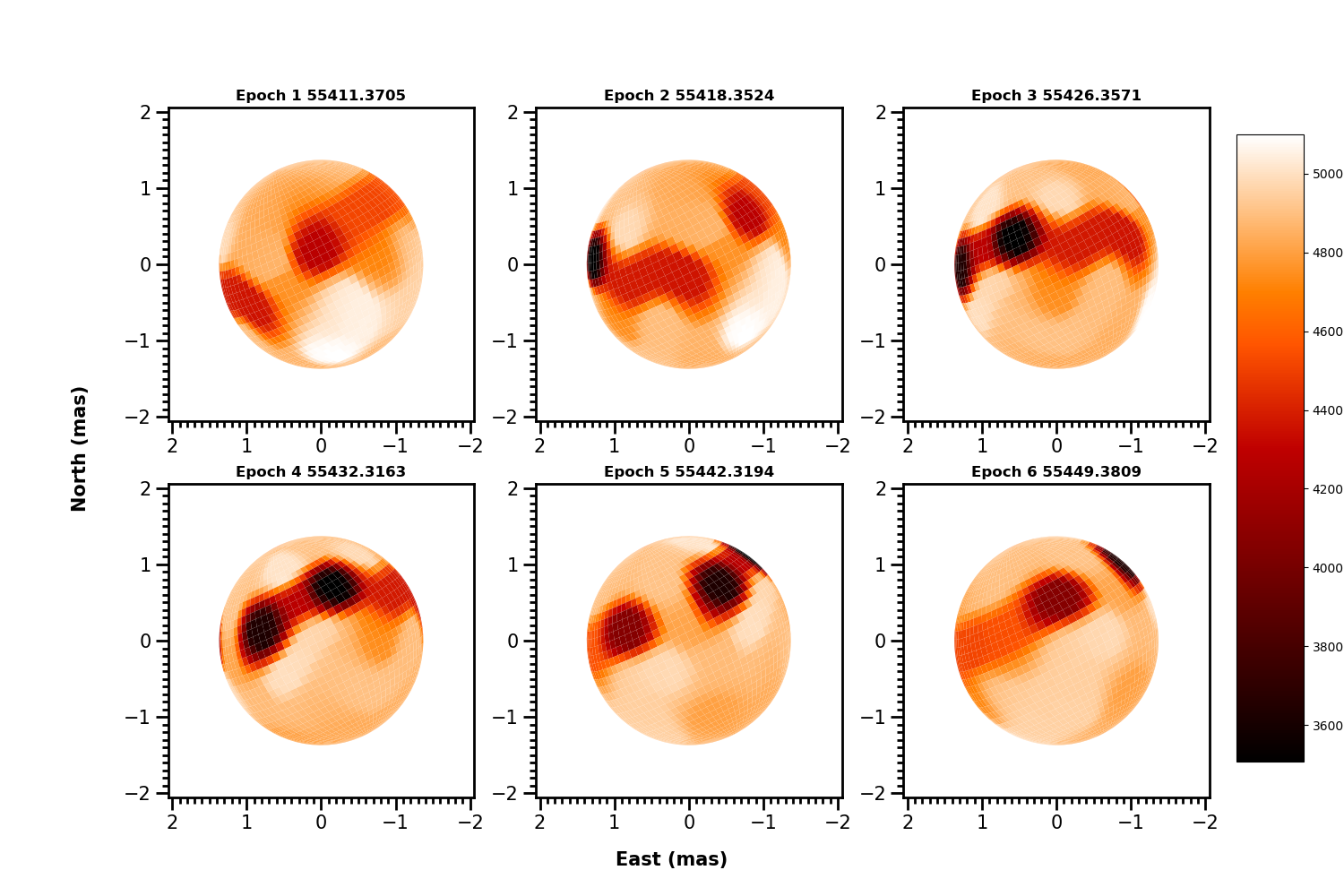}
    \includegraphics[width=0.5\textwidth]{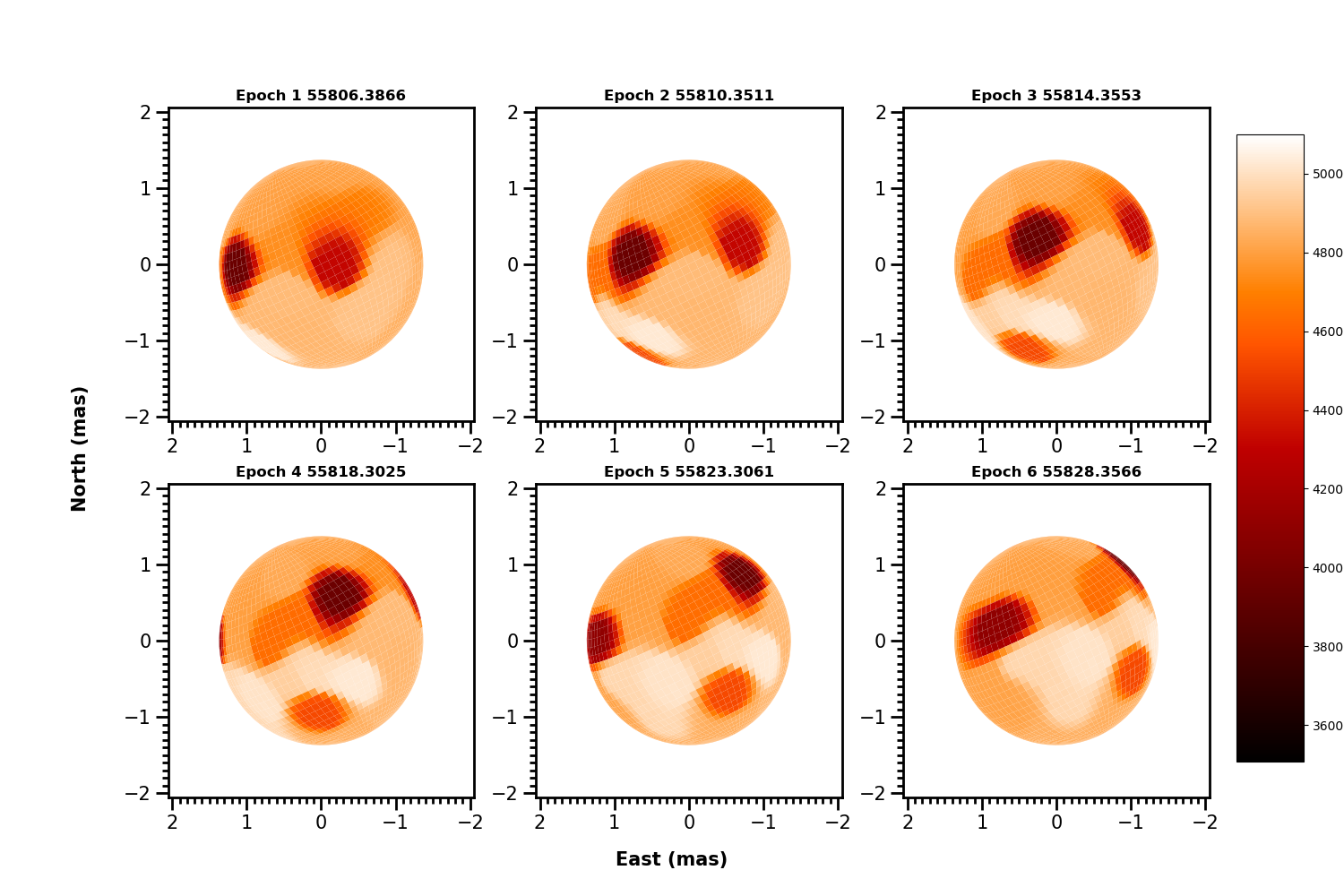}
    \caption{We show temperature maps (in Kelvin) of $\lambda$~Andromedae for the 2010 epoch (left) and 2011 epoch (right) using our \texttt{ROTIR} code. Here, we note that our 2010 temperature map panels do not reflect all 11 nights of data but only show a subset of 6 nights. The nights for the 2010 temperature map panels are chosen by only selecting one of two consecutive observational nights and having the next temperature map panel be separated by at least 6 nights (i.e., 2010-Aug-03, 2010-Aug-10, 2010-Aug-18, 2010-Aug-24, 2010-Sep-03, 2010-Sep-10). \label{fig:temperature_map}}
\end{figure*}

Once we have determined that the spherical geometrical setup is the most optimal for $\lambda$~And and choose the most optimal regularization weight, we are now set for calculating the best fit for the temperature map. We present the resulting Mollweide maps of $\lambda$~And for both epochs in Figure \ref{fig:Mollweide_map}. These maps reflect no time variability and assume that $\lambda$~And is undergoing solid-body rotation. A better representation of the temperature maps are shown in Figure \ref{fig:temperature_map} for each given night in 2010 and 2011.

A first look at the temperature maps between the 2010 and 2011 epochs shows a few interesting characteristics about $\lambda$ And's surface. Comparing the two temperature maps show notable similarities for two spots in the northern hemisphere between the two epochs (i.e., the spot around $20^\circ$ latitude and $\sim100^\circ$ longitude, and the spot around $0^\circ$ latitude and $170^\circ$ in both epochs). There are two other notable spots that either disappear or appear from one epoch to the next. The spot in the 2010 epoch around $30^\circ$ latitude and $150^\circ$ longitude seems to has disappeared within the 2011 epoch. A spot seems to be forming within the 2011 map in the southern hemisphere around $-40^\circ$ latitude and $50^\circ$ longitude with hints of its emergence with similar place in the 2010 epoch. We note that the spot in the 2010 epoch around $15^\circ$ latitude and $-90^\circ$ does not appear in the 2011 epoch. This is most likely due to missing rotational phase coverage in the 2011 data set.

\subsection{Refinement of physical parameters}
\begin{figure*}[ht!]
    \includegraphics[width=0.24\textwidth]{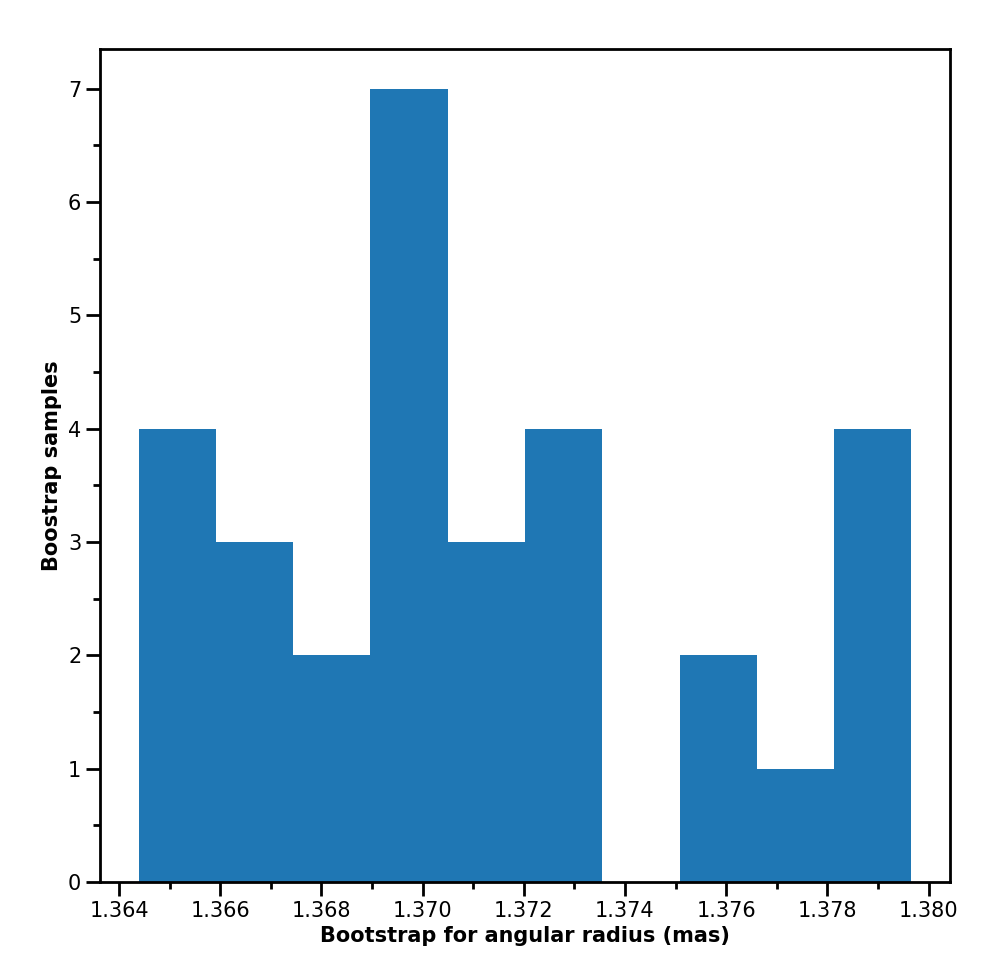}
    \includegraphics[width=0.24\textwidth]{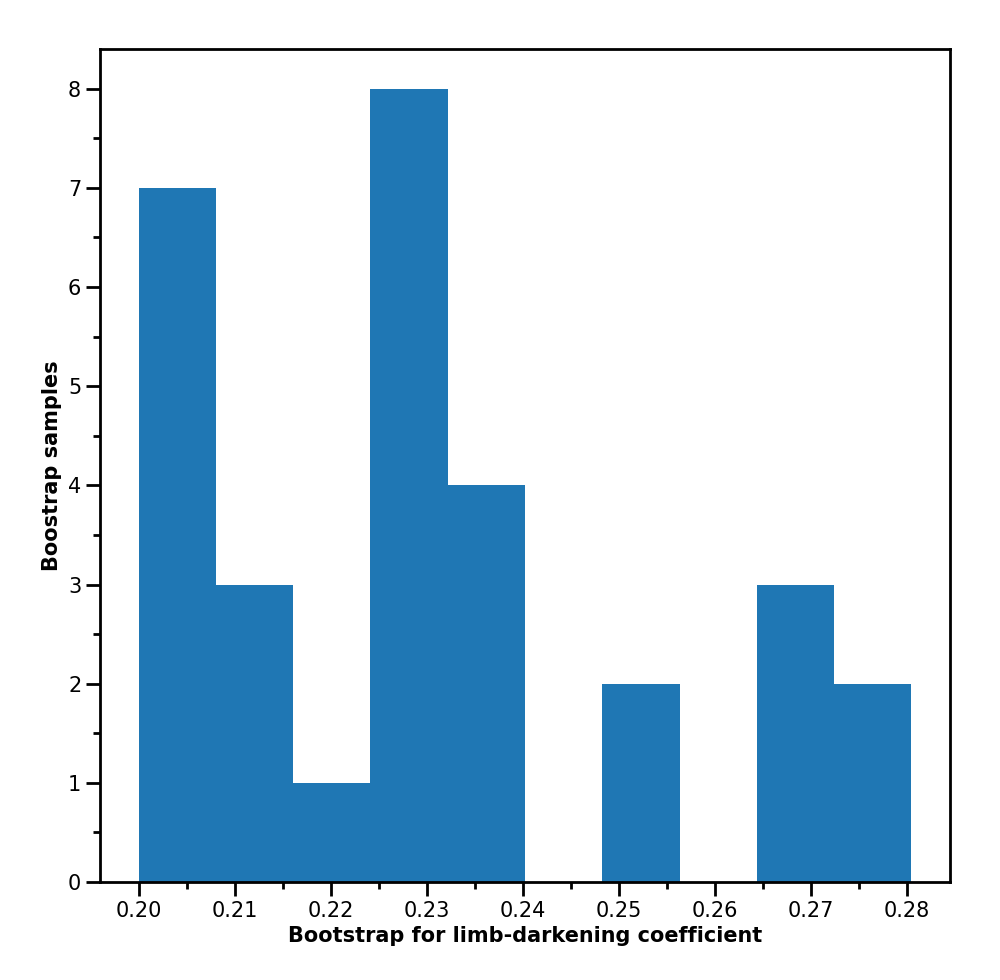}
    \includegraphics[width=0.24\textwidth]{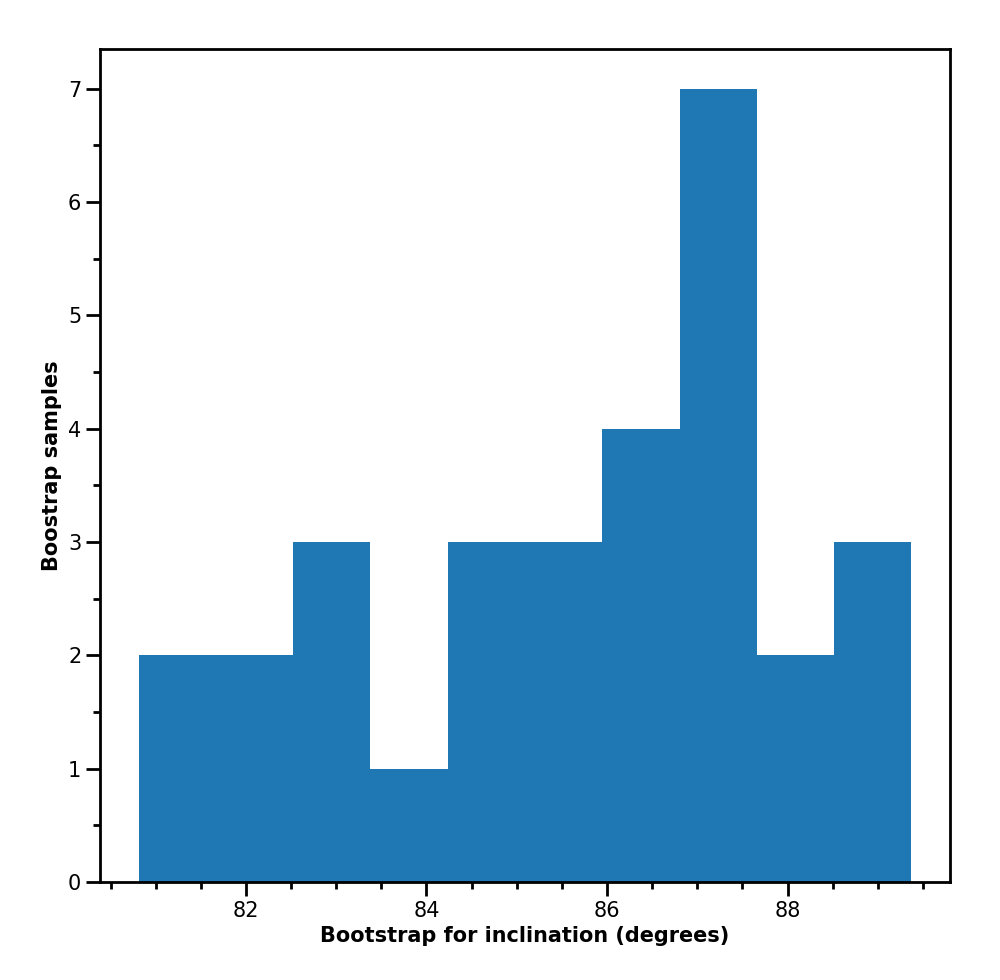}
    \includegraphics[width=0.24\textwidth]{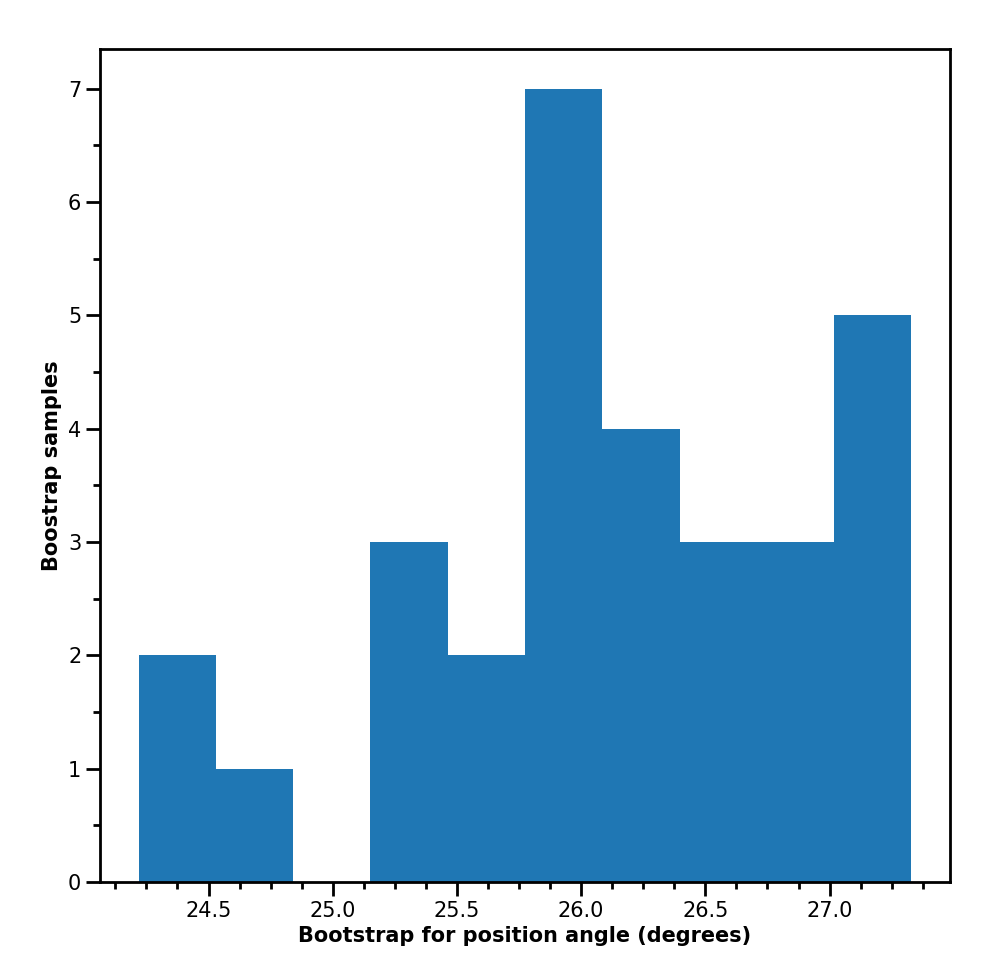}
    \caption{Here we show the results of using the bootstrap method varying angular radius, the limb-darkening coefficient, inclination, and position angle. We use 50 bootstraps in order to calculate the final parameters of $\lambda$~Andromedae and bin them into 10 different bins. The x-axis here shows the range of the parameters from all the bootstraps and the y-axis show the number of bootstraps within each bin. While we plot calculated values for each bootstrap, we note that that the full range for each parameter are the following: [1.35, 1.39] for angular radius, [0.2, 0.3] for the limb-darkening coefficient, [70.0, 90.0] for inclination, and [20.0, 30.0] for position angle. The final parameters are calculated from taking the average of each respective parameter with their associated errors calculated from the standard deviation of the bootstrap results. \label{fig:bootstrap_primary}}
\end{figure*}

After finding the best model from \texttt{SIMTOI}, we use the parameters from the 4-spot model based on the 2011 data and use the bootstrap method. We apply the bootstrap method in order to find the final parameters and errors for the primary component of $\lambda$~And. We use 50 bootstrap iterations to solve for only four parameters: angular radius, the limb-darkening coefficient, inclination, and position angle. We choose to leave the rotation period of the primary fixed throughout this bootstrap because there is a degeneracy towards lower rotation periods. We believe that this is due to the fitting algorithm in \texttt{ROTIR} choosing the difference between the first and last observing date (in a given epoch) for the period instead of the true rotational period. Our bootstrap is dependent on the NLopt package \citep{johnson:2008} and Nelder-Mead Simplex method \citep{nelder_mead:1965,box:1965,richardson_kuester:1973} within NLopt for obtaining our final parameters with their corresponding errors. We restrict lower and upper bounds within NLopt for these four parameters as follows: [1.35, 1.39] for angular radius, [0.2, 0.3] for the limb-darkening coefficient, [70.0, 90.0] for inclination, and [20.0, 30.0] for position angle. The final values for each variable parameter are chosen by averaging over all bootstraps and their associated errors are calculated though their standard deviation. Our 50 bootstraps do not show a Gaussian distribution, but we are prevented from running a large number of bootstraps due to computation time. We indicate that parameters do not deviate too largely from their mean values. It is likely that doing more bootstraps will slightly increase the error bars but not in a significant manner. We show the results of our bootstrap values in Figure \ref{fig:bootstrap_primary}.

\subsection{Images of \texorpdfstring{$\lambda$}{lambda}~And} \label{sec:rotir_params}
\begin{figure*}[ht!]
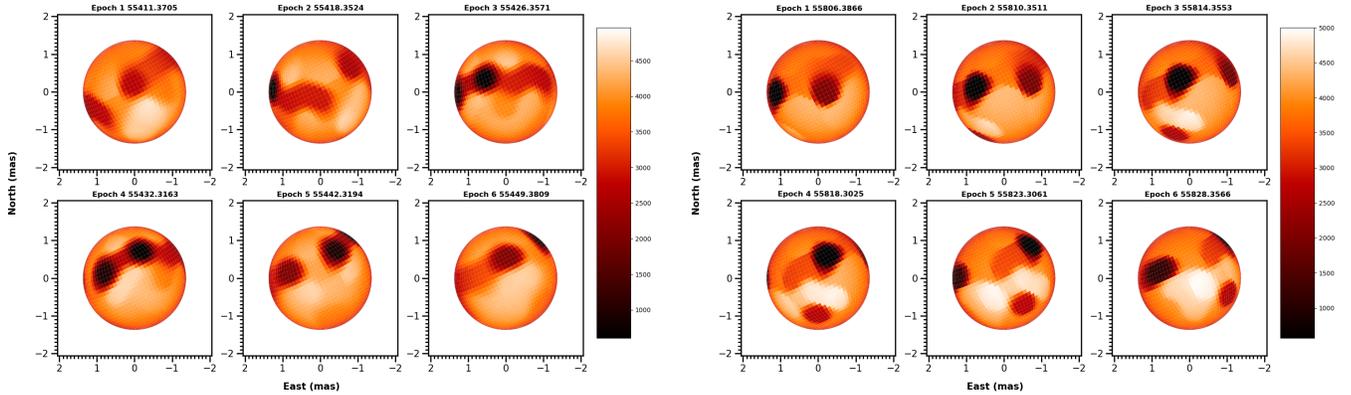

    \includegraphics[width=0.5\textwidth]{figures/intensity_map2010_static_cmap_final_v2.png}
    \includegraphics[width=0.5\textwidth]{figures/intensity_map2011_static_cmap_final_v2.png}
    \caption{We show relative intensity maps (in arbitrary units) of $\lambda$~Andromedae for the 2010 epoch (left) and 2011 epoch (right) using our \texttt{ROTIR} code. Here, we note that our 2010 intensity map panels do not reflect all 11 nights of data but only show a subset of 6 nights. The nights for the 2010 intensity map panels are chosen by only selecting one of two consecutive observational nights and having the next intensity map panel be separated by at least 6 nights (i.e., 2010-Aug-03, 2010-Aug-10, 2010-Aug-18, 2010-Aug-24, 2010-Sep-03, 2010-Sep-10). All images here for both the 2010 and 2011 epochs reflect the same parameters that are listed in Table \ref{tab:lambda_Andparams}.
    \label{fig:intensity_map}}
\end{figure*}

Temperature maps are not indicative of what is actually represented from observations. In order to present an image, we include to use a power law for limb-darkening \citep{hestroffer:1997} and multiply it by the cells of the temperature maps that are visible to the observer. We use the limb-darkening coefficient from our bootstrap to present the images in Figure \ref{fig:intensity_map} and present the physical parameters for the primary star in $\lambda$~And using the parameters from our bootstrap in Table \ref{tab:lambda_Andparams}.

\begin{deluxetable*}{lCcCl}
\tablenum{4}
\tablecaption{Final $\lambda$ Andromedae Parameters for the Primary \label{tab:lambda_Andparams}}
\tablecolumns{5}
\tablewidth{0pt}
\tablehead{
\colhead{Observed Parameters} &
\colhead{Value} &
\colhead{Source} &
\colhead{Values from literature} & 
\colhead{Literature reference}
}
\startdata
$R_{\star}$ (mas) & 1.371 \pm 0.005 & This work & 1.379 \pm 0.025 & \cite{parks:2021} \\ 
Limb-darkening coefficient & 0.231 \pm 0.024 & This work & 0.229 \pm 0.111 & \cite{parks:2021} \\ 
Inclination (deg) & 85.63 \pm 2.32 & This work & 70.35 \pm 6.7\tablenotemark{d} & \cite{parks:2021} \\ 
Position Angle (deg) & 26.09 \pm 0.82 & This work & 21.6 \pm 7.5\tablenotemark{d} & \cite{parks:2021} \\ 
Rotation Period (days) & 54.2 & This work & 56.9 \pm 8.8\tablenotemark{d} & \cite{parks:2021} \\
\hline
Physical parameters & & & \\
\hline
$R_{\star}$ ($R_{\odot}$) & 7.787 \pm 0.053 & This work\tablenotemark{a} & 7.831^{+0.067}_{-0.065} & \cite{parks:2021}\\ 
$T_{\mathrm{eff}}$ (K) & 4800 \pm 100 & \cite{drake:2011} & - & $-$ \\
log $g$ & 2.75\pm 0.25 & \cite{drake:2011} & - & $-$ \\
$M_{\star}$ ($M_{\odot}$) & 1.24 \pm 0.72 & This work\tablenotemark{b} & 1.3^{+1.0}_{-0.6} & \cite{drake:2011} \\
log $L/L_{\odot}$ & 1.46 \pm 0.04 & This work\tablenotemark{c} & 1.37 \pm 0.04 & \cite{drake:2011} \\
distance (pc) & 26.41 \pm 0.15 & \cite{vanleeuwen:2007} & - & $-$ \\
\enddata
\tablenotetext{a}{Based on the angular radius from this work and the distance from \cite{vanleeuwen:2007}.}
\tablenotetext{b}{Based on the physical radius from this work and the log $g$ from \cite{drake:2011}.}
\tablenotetext{c}{Based on the physical radius from this work and the effective temperature from \cite{drake:2011}.}
\tablenotetext{d}{Since \cite{parks:2021} had multiple values reported for the same parameter, we show the averages of the respective parameter here.}
\tablecomments{The observed parameters were optimized through a bootstrap approach with the exception of the rotation period, which was fixed. We take our fixed rotation period parameter directly from the best model in \texttt{SIMTOI}.}
\end{deluxetable*}

\section{Comparisons to previous work} \label{sec:compare_previous}

\subsection{\texttt{SURFING} vs \texttt{ROTIR} imaging} \label{sec:surfing}
\begin{figure*}[ht!]
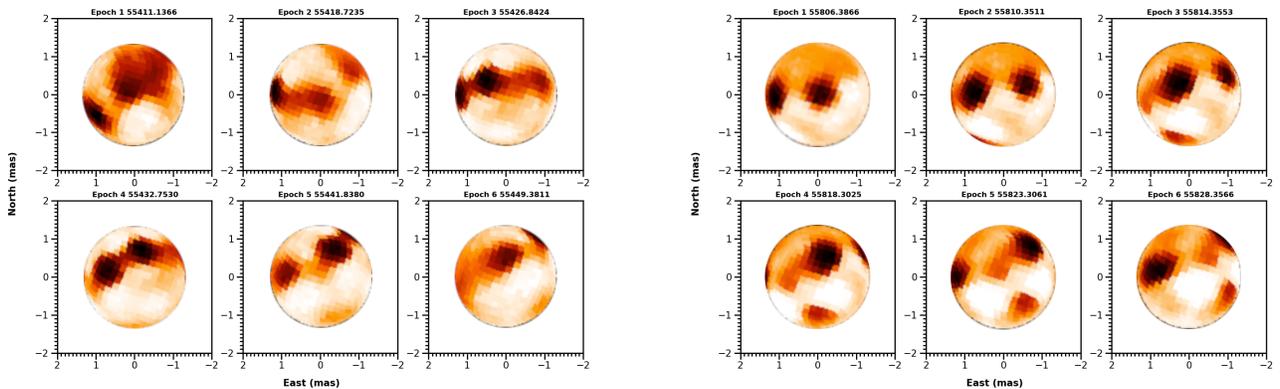

    \includegraphics[width=0.5\textwidth]{figures/2010_lamAnd_SURFING_v4.png}
    \includegraphics[width=0.5\textwidth]{figures/2011_lamAnd_SURFING_v4.png}
    \caption{Temperature maps of $\lambda$~Andromedae in 2010 (left) and 2011 (right) using \texttt{SURFING} code. The 2010 temperature maps (in similar Kelvin scale to Figure \ref{fig:temperature_map}) were made by using two different consecutive nights and merging the data as one night. We find that this does not largely affect the results of the imaging since the rotation made from two consecutive nights only span $\sim$2\% of the rotation period. \label{fig:SURFING_tempmap}}
\end{figure*}

Here we compare images made independently from \texttt{ROTIR} to another image reconstruction code called \texttt{SURFace imagING} (\texttt{SURFING}) in Figure \ref{fig:SURFING_tempmap}. \texttt{SURFING} is a Monte Carlo based imaging code written in IDL specifically written for imaging spheroids \citep[see][]{roettenbacher:2016a}. Overall, there is a good agreement between the two imaging methods. Since we are only focusing on the imaging comparison aspect for these two codes, we see that the spot locations and contrast between the two are very similar, with a few minor differences, as shown in Figures \ref{fig:temperature_map} and \ref{fig:SURFING_tempmap}.

\subsection{Comparison to Parks et al.}
The results of this work largely agree to those of \cite{parks:2021} with exception of the inclination of $\lambda$~And being the only disagreement. \cite{parks:2021} used a combination of a genetic algorithm \citep{charbonneau:1995} and the Nelder-Mead Simplex method \citep{nelder_mead:1965, box:1965, richardson_kuester:1973}, in order to make individual models for each night of data. Each surface model calculates an angular diameter, limb-darkening coefficient based on the power law, a starspot covering factor, starspot latitude, starspot longitude, and starspot intensity ratio for $\lambda$ And. Once all the models were made, \cite{parks:2021} traced each starspot on the surface for each epoch. Ellipse fits to starspot positions were calculated, and an average computed position angle and inclination angle were made from these ellipse fits for each year.

\cite{parks:2021} reported that the inclination of primary from their 2010 and 2011 data is $75\pm5.0^{\circ}$ and $66.4\pm8.0^{\circ}$, for each respective year, giving an overall average of $70.35\pm6.7^{\circ}$ while we report an inclination of $85.63\pm2.32^{\circ}$. We believe that our calculations from this work are accurate for several reasons. The initial \texttt{SIMTOI} calculations were done with a global search with no restrictions in parameter space, including inclination. The resulting parameters obtain from \texttt{SIMTOI} were then used in \texttt{ROTIR} with a sufficient range that included the inclination value from \cite{parks:2021}. If the value for our inclination were incorrect and actually leaned towards this previous value, the resulting bootstrap method would have reflected it by converging on the lower bounds of our parameter space using our bootstraps. In addition, the work by \cite{parks:2021} relied on independent models for each night and tied them together to form an analysis while we use the all the data of each epoch collectively to form one image.

\section{Beyond solid rotation imaging} \label{sec:solid_rot}
\subsection{Simulating differential rotation}
In our Figures \ref{fig:Mollweide_map} - \ref{fig:SURFING_tempmap} using \texttt{SIMTOI}/\texttt{ROTIR}, \texttt{SURFING}, and in \cite{parks:2021}, all imaging has been performed assuming that the star is rotating as a solid body however, we attempt to estimate differential rotation through our data. \cite{henry:1995b} studied photometry of $\lambda$~And over 14 years and found evidence of shear across the surface. In order to see if we are able to detect any differential rotation with our interferometric data, we simulate starspots on a star with a low differential rotation coefficient and a low temperature gradient on the surface, and have the spot move across a few days with the same period as $\lambda$~And. Then we do a cross-correlation for each latitude band on the star and see if there is any deviation from zero.

\begin{figure*}
    \includegraphics[width=0.333\textwidth]{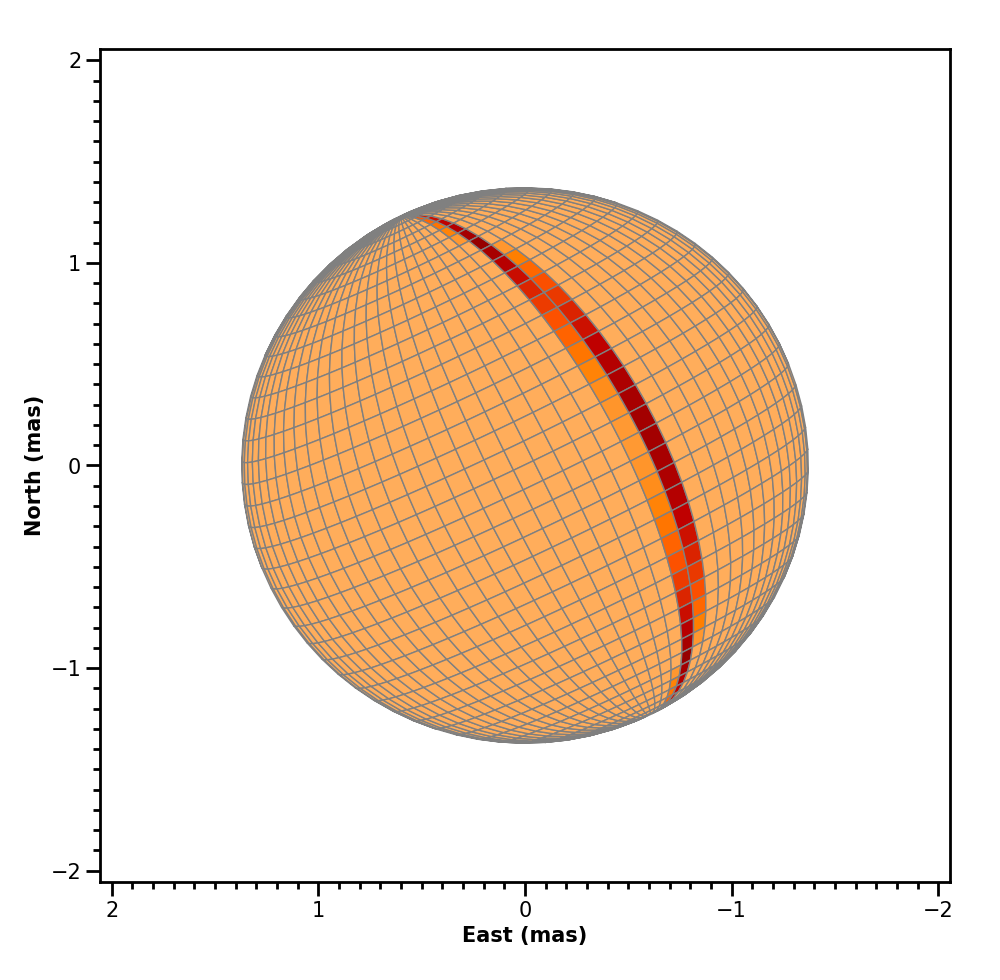}
    \includegraphics[width=0.333\textwidth]{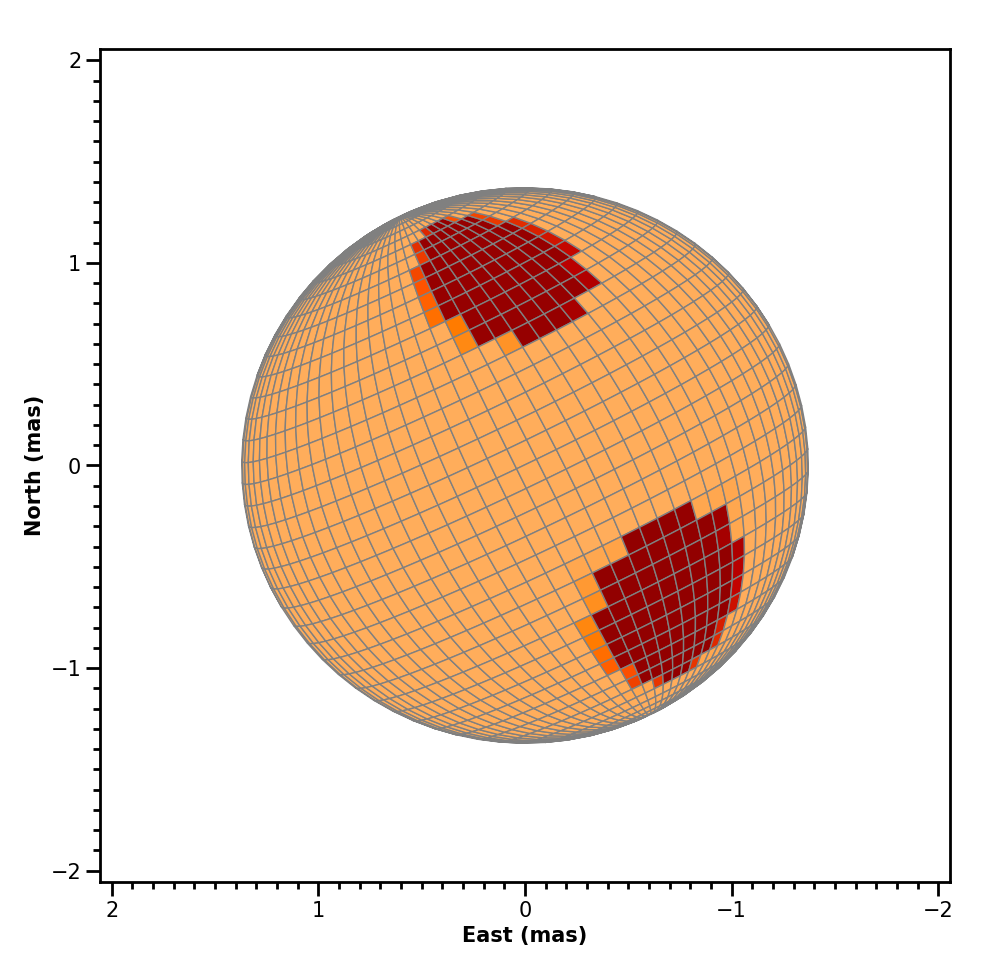}
    \includegraphics[width=0.333\textwidth]{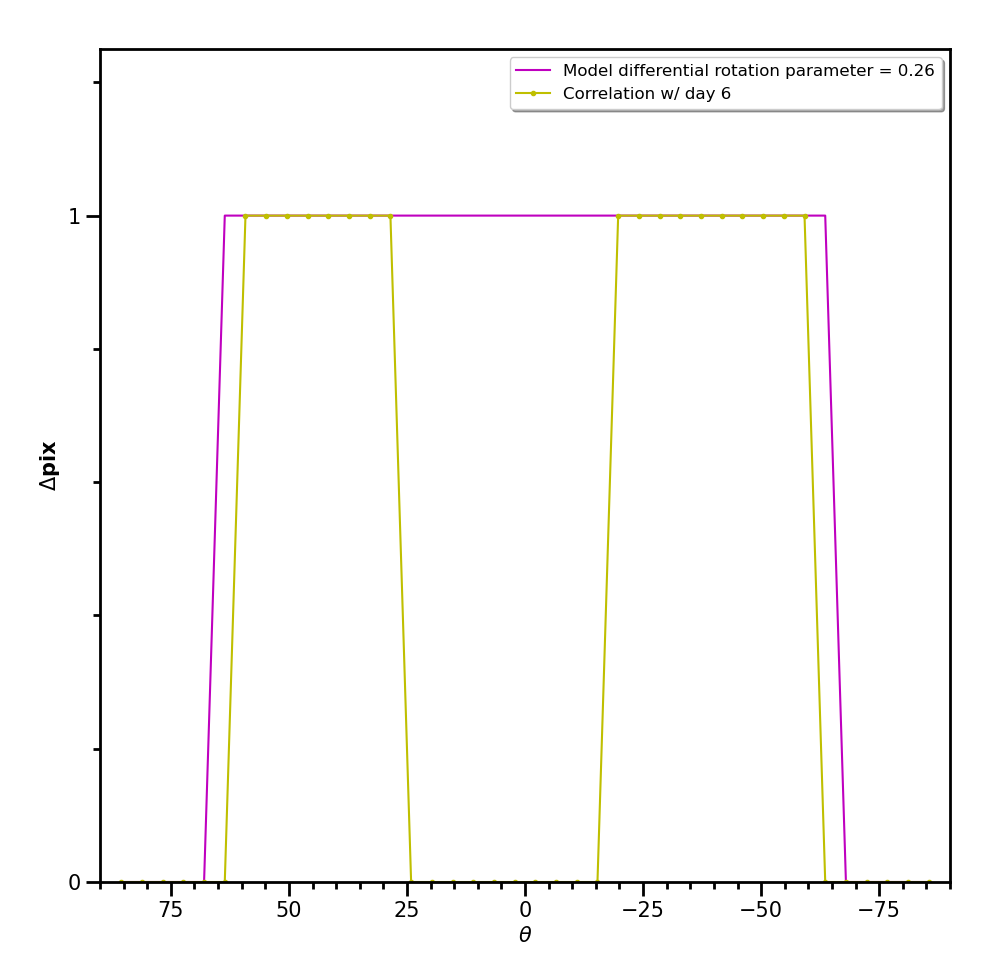}
    \caption{We show simulations of differential rotation by doing a correlation using the unrealistic starspot among a longitudinal band (left) and two starspots (middle) of a fake star with the same parameters of $\lambda$~Andromedae within the 2011 epoch (with the exception of the temperature map). The differential rotation coefficient we use here is $\Delta \Omega$ of 0.26 from \cite{henry:1995b}. The plot (right) shows the number of pixels that have shifted in respect to the longitude after subtracting off the total shift of a spot for a given latitude. The pink line at this coefficient represents the unrealistic starspot change in pixels while the yellow line shows the two starspots change in pixels as a function of the longitude. We choose to compare the first and last observations within the 2011 epoch to show the maximum amount of correlation.
    \label{fig:diff_rot_sim}}
\end{figure*}

Our simulations show two different scenarios. The first simulation presents a highly unrealistic starspot that are two pixels wide in longitude and spanning throughout all latitude from pole to pole. Our second simulation shows two circular starspots that are 5 pixels in radius at $+45^{\circ}$ and $-45^{\circ}$ latitude (in respect to the equator) and at $135^{\circ}$ longitude. We presents our simulations of a simple star with similar parameters as $\lambda$~And using differential rotation coefficient from \cite{henry:1995b} of k = 0.04, which corresponds to differential angular velocity ($\Delta \Omega$) of 0.26, in Figure \ref{fig:diff_rot_sim}. 

\subsection{Testing differential rotation on \texorpdfstring{$\lambda$}{lambda} And}
\begin{figure}
    \includegraphics[width=0.45\textwidth]{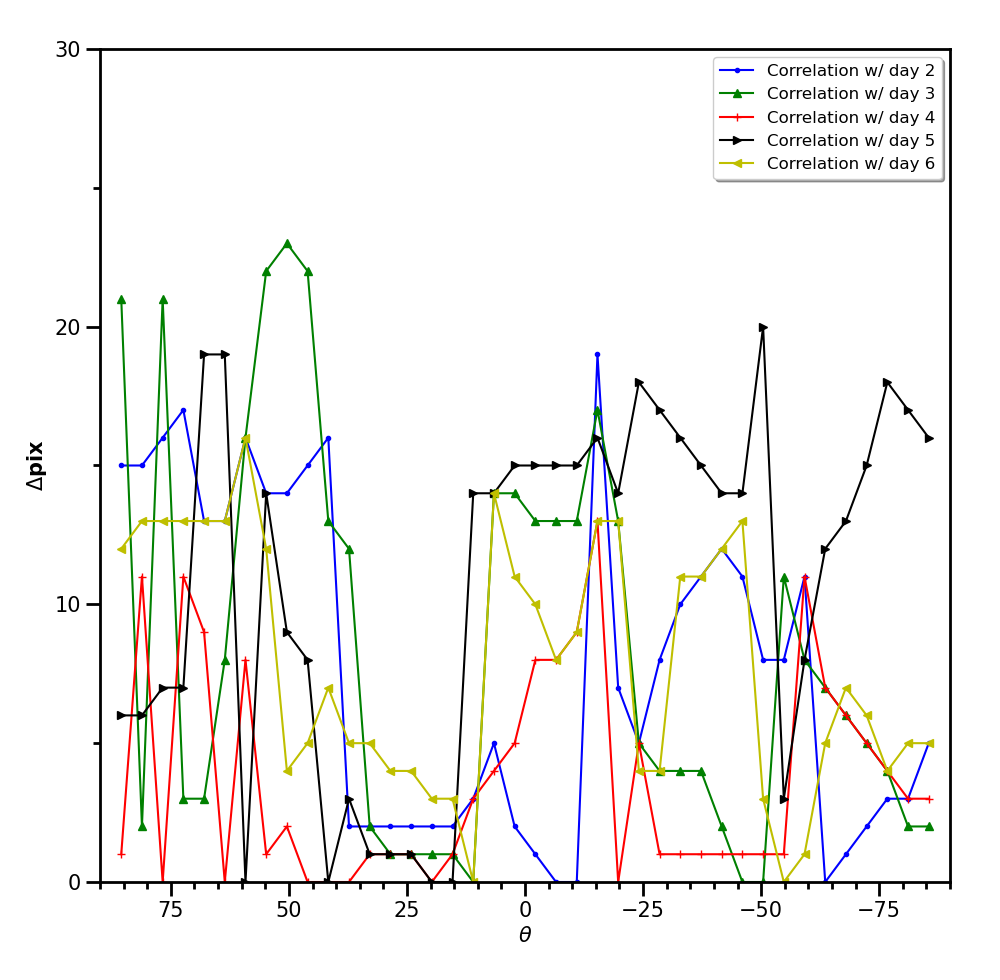}
    \includegraphics[width=0.45\textwidth]{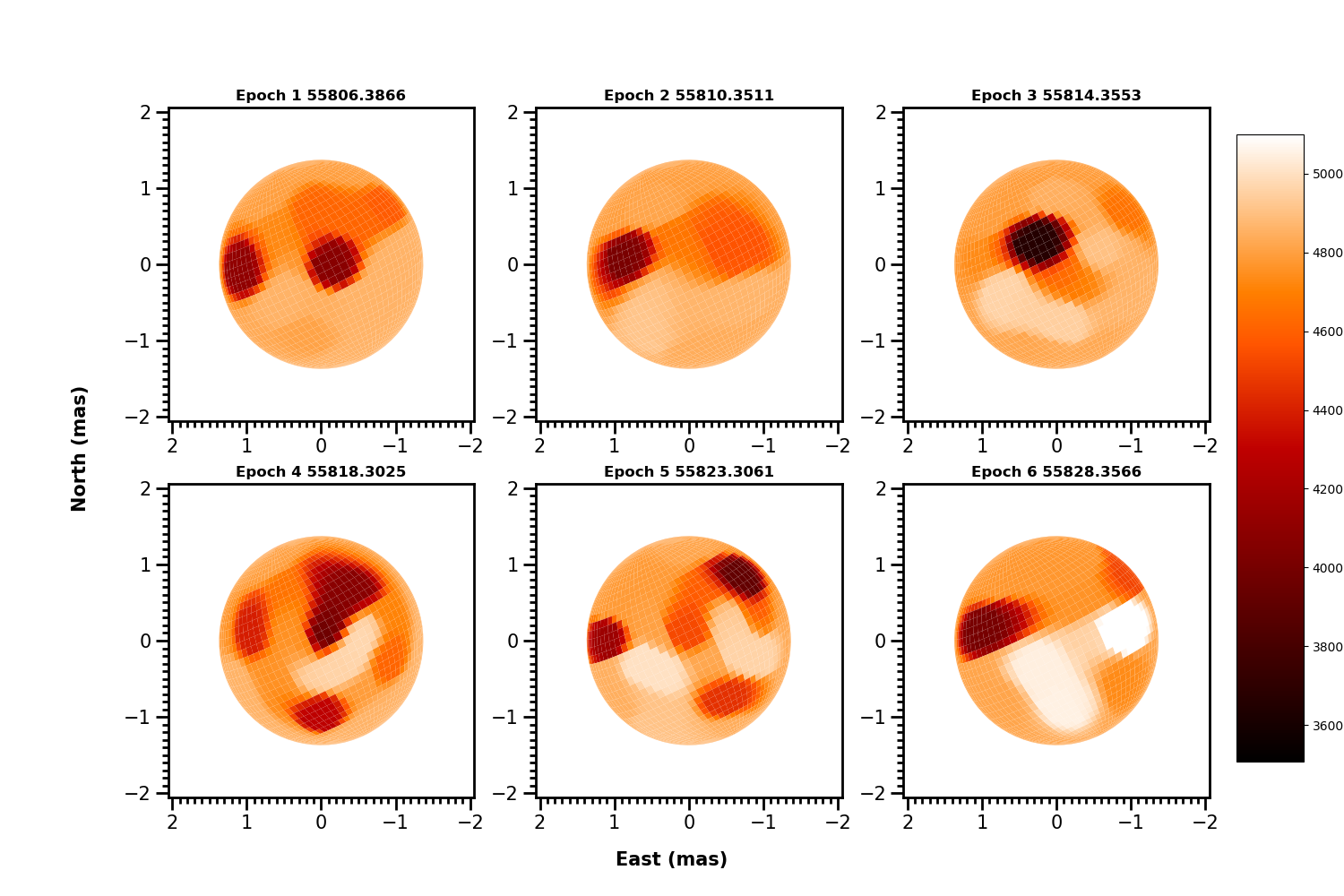}
    \caption{The plot shown here (top) is similar to that of Figure \ref{fig:diff_rot_sim} but with the actual $\lambda$~Andromedae data. The different symbols denote correlations of temperature maps compared to the first observation of $\lambda$~Andromedae in the 2011 epoch. The individual temperature maps (in Kelvin) for each observation date in 2011 (bottom) were constructed using the original temperature map from Figure \ref{fig:Mollweide_map}. These maps reflect the difficulty in searching for shear at the one pixel level since each map is slightly different compared to the previous observation and results in no visible correlation.
    \label{fig:diff_rot_lamAnd}}
\end{figure}

We apply the same cross-correlation method for the 2011 data set and calculate the deviations. We find that we are unable to detect any differential rotation with our data due to three reasons. First, our data does not span an entire rotation, therefore we are not able to compare the same spots from the previous rotation. Second, $\lambda$~And is a very slow rotator so we do not have enough resolution to detect any small amounts of differential rotation, if differential rotation truly exists on $\lambda$~And. In fact, the large scale magnetic spots on $\lambda$~And may not be able to be used to measure any real surface differential rotation based on its dynamo. \cite{korhonen:2011} states that surface differential rotation can only be recovered by observing the spot motion of small spots, unlike $\lambda$~And's large scale magnetic spot structure. Third, the amount of square visibilities and closure phases for each observation are sparse for most observations. Since the goal is to detect any shear as evidence for differential rotation, we reconstruct an individual temperature map for each observation date from the 2011 epoch but initialize with the temperature map obtained from Figure \ref{fig:Mollweide_map}. We show our results in Figure \ref{fig:diff_rot_lamAnd}. 

\section{Beyond Imaging the Primary} \label{sec:binary_search}
\subsection{Updated Orbital Parameters and Secondary Parameters}
Using the updated parameters from the primary star in $\lambda$~And in this work, the mass ratio from \cite{donati:1995} and Kepler's Third Law, we are now able to calculate the mass of the secondary and the semi-major axis of the binary system. We calculate that the mass for the companion is $0.15^{+0.09}_{-0.05}~M_{\odot}$ and with corresponding semi-major axis is $6.12$ mas for the system.
\subsection{The Search for the Secondary}
We begin our search for the companion by obtaining an estimate on the luminosity ratio and angular size of the secondary to narrow down our search. For the luminosity ratio, we used a mass-luminosity relation for each corresponding star in our system ($L_2/L_1 = 0.23(M_2^{2.3}/M_{1}^{4})$) and calculated to be approximately $L_2/L_1 = 0.00121$. If we assume that the H-band flux ratio is the same as the luminosity ratio of the two stars and if we use the H-band magnitude of the primary 1.40~mag \citep{ducati:2002}, this would correspond to an estimated H-band magnitude of 8.7~mag for the secondary. This is slightly beyond MIRC's magnitude limit and not likely to be detected, however we still investigate the possibility of detection. In order to calculate the estimated angular size of the secondary, we first calculate the physical size by using the mass-radius relation ($R = 0.0753 + 0.7009M + 0.2356M^2$) developed by \cite{maldonado:2015} for low-mass stars. Given that the calculated physical radius is $0.19~R_{\odot}$, we find that the estimated angular radius is approximately $0.03$ mas.

Now that we have an estimation of the angular size and flux ratio, we perform a grid search in right ascension and declination over a 10 mas distance from the primary star for every night in the 2011 epoch. This approach is similar the methods used in \cite{baron:2012} and CANDID \citep{gallenne:2015} with the difference that the primary is using the model visibilities obtained during image reconstruction. We model binary visibilities and vary both the brightness ratio and the angular radius for the secondary using NLopt for each section of the grid. We restrict the parameter space for the angular radius to [0.0, 1.0] mas while restricting the flux ratio (secondary/primary flux) for the system from [0.0, 0.2]. 

While we do find that the average flux ratio using the 2011 data set of $0.00213 \pm 0.00116$ is within the theoretical estimated value, we find two major reasons for believing that we were not able to find the secondary companion. First, the average angular radius found by using the 2011 data is $0.602 \pm 0.356$ mas, largely inconsistent with our estimation using mass-radius relation for low-mass stars. Our errors for both the flux ratio of the system and the angular size of the secondary were calculated by taking the standard deviation of every night's grid search result from the 2011 epoch. The values of angular radius for an individual night were also seen to hit a boundary condition (either 0 mas or 1 mas), thus assessing that the calculated values are incorrect. Second, the best fit right ascension and declination positions for each night in the 2011 data set were positioned in a random assortment on the grid space with no clear indication of a circular or elliptical orbit. 

Another reason that we may not be able to find the secondary for $\lambda$~And could be due to lack of ($u,v$) coverage for each individual night in the 2011 epoch data set. For this reason, we proceed to not use the 2010 data set to find the secondary as those observations were taken with two different sets of 4T observations in a given night and as a result do not provide better ($u,v$) coverage compared to the 2011 data set.

\section{Conclusion and Future Prospects} \label{sec:conclusion}
In this paper, we do interferometric modeling and imaging on $\lambda$~And for the 2010 and 2011 epochs. First, we use \texttt{SIMTOI} in order to find which model is most probable for finding the best parameters. Then we use the parameters from \texttt{SIMTOI} and use them for imaging in \texttt{ROTIR}. Using the parameters from the best \texttt{SIMTOI} model as a starting point, we apply the bootstrap method to get the final physical parameters for $\lambda$~And. We find that our images from \texttt{ROTIR} fairly agree with the images produced to the other image reconstruction code, \texttt{SURFING}, and our physical parameters are also fairly consistent of previous works with the exception of the inclination.

Images from both codes show that the spots on $\lambda$~And from both epochs seem to favor certain latitudes and are mostly concentrated in the northern hemisphere. For both the 2010 and 2011 epochs, we find that most of the spots are centered around $+20^{\circ}$ latitude. These spot concentrations to a certain latitude are consistent with the interferometric images shown in \cite{roettenbacher:2016a} of $\zeta$ Andromedae, another RS CVn variable. The absence of symmetrical spots on active latitudes as observed on the Sun is evidence that $\lambda$~And may not have a solar-like dynamo.

Finally, once we produce static images of the primary star in the system, we test to see if we find any evidence for differential rotation and detect the secondary companion. We start with a simulation of differential rotation and compare those results to the 2011 interferometric data set. Our results remain inconclusive as we cannot detect any sheer within the 2011 data set largely due to $\lambda$~And being a slow rotator. In our search for the companion, we do a grid search by fitting various models for the companion (i.e., varying the angular radius of the secondary and flux ratio of the system). While the flux ratio was consistent with the approximated value, the angular radius was largely inconsistent with our estimated calculation therefore concluding that we were unable to detect the secondary. 

Our \texttt{ROTIR} code is not just limited to interferometric imaging but is also capable of light-curve inversion. Our future work will plan on using the multi-band photometry in \cite{parks:2021} and compare those resulting images with the interferometric images from this work. Our plans also include using the photometric data as a bridge for the 2010 and 2011 interferometric epochs in order to detail how $\lambda$~And is evolving over the course of a year. We are currently implementing additional numerical techniques to \texttt{ROTIR} \citep{abbott:2021} in order to improve light-curve inversion quality with the use of Alternating Direction Method of Multipliers \citep{chan:2011}. Finally, we have future plans to implement Doppler imaging and Zeeman-Doppler imaging into \texttt{ROTIR}.

\acknowledgments

The authors thank the anonymous referee for the insightful comments that improved the quality of the manuscript. AOM and FRB acknowledge support from NSF Grant No. AST-1616483 and AST-1814777. RMR acknowledges support from the Yale Center for Astronomy and Astrophysics (YCAA) Prize Postdoctoral Fellowship.

This work is based upon observations obtained with the Georgia State University Center for High Angular Resolution Astronomy Array at Mount Wilson Observatory. The CHARA Array is supported by the National Science Foundation under Grant No. AST-1636624 and AST-1715788. Institutional support has been provided from the GSU College of Arts and Sciences and the GSU Office of the Vice President for Research and Economic Development. Any opinions, findings, and conclusions or recommendations expressed in this material are those of the author(s) and do not necessarily reflect the views of the National Science Foundation.

\facilities{The Center of High Angular Resolution Astronomy Array}

\software{\texttt{SIMTOI}, \texttt{OITOOLS}, \texttt{ROTIR}}
\newline

\bibliographystyle{aasjournal}
 \newcommand{\noop}[1]{}


\end{document}